\documentclass[sigconf]{acmart}

\AtBeginDocument{%
  \providecommand\BibTeX{{%
    \normalfont B\kern-0.5em{\scshape i\kern-0.25em b}\kern-0.8em\TeX}}}

\acmConference[CCS' 19]{CCS'19}
\acmBooktitle{XXX}

\renewcommand\footnotetextcopyrightpermission[1]{} 
\usepackage{enumerate}
\usepackage{tikz}
\usepackage{amsmath}
\usepackage{hyperref}

\usepackage{color} 
\usepackage{multirow}
\usepackage{threeparttable}
\usepackage{amssymb}
\newcommand*\dash{\unskip\kern.16667em---\penalty\exhyphenpenalty
        \hskip.16667em\relax
}

 {\makeatletter
  \gdef\xxxmark{%
    \expandafter\ifx\csname @mpargs\endcsname\relax 
      \expandafter\ifx\csname @captype\endcsname\relax 
        \marginpar{\textcolor{red}{xxx}}
      \else
        \textcolor{red}{xxx~}
      \fi
    \else
      \textcolor{red}{xxx~}
    \fi}
  \gdef\xxx{\@ifnextchar[\xxx@lab\xxx@nolab}
  \long\gdef\xxx@lab[#1]#2{{\bfseries [\xxxmark \textcolor{red}{#2}
  ---{\scshape #1}]}}
  \long\gdef\xxx@nolab#1{{\bfseries [\xxxmark \textcolor{red}{#1}]}}
 }

\ifx \debug
    \newcommand{\yzj}[1]{{}}
    \newcommand{\yz}[1]{{}}
    \newcommand{\wenhui}[1]{{}}
\else
    \newcommand{\yzj}[1]{{\bf\textcolor{blue}{[Yzj: #1 ]}}}
    \newcommand{\wenhui}[1]{{\bf\textcolor{red}{[WH: #1 ]}}}
    \newcommand{\yz}[1]{{\bf\textcolor{cyan}{[YZ: #1 ]}}}
\fi

\usepackage{url}
\makeatletter
\def\UrlAlphabet{%
      \do\a\do\b\do\c\do\d\do\e\do\f\do\g\do\h\do\i\do\j%
      \do\k\do\l\do\m\do\n\do\o\do\p\do\q\do\r\do\s\do\t%
      \do\u\do\v\do\w\do\x\do\y\do\z\do\A\do\B\do\C\do\D%
      \do\E\do\F\do\G\do\H\do\I\do\J\do\K\do\L\do\M\do\N%
      \do\O\do\P\do\Q\do\R\do\S\do\T\do\U\do\V\do\W\do\X%
      \do\Y\do\Z}
\def\UrlDigits{\do\1\do\2\do\3\do\4\do\5\do\6\do\7\do\8\do\9\do\0}
\g@addto@macro{\UrlBreaks}{\UrlOrds}
\g@addto@macro{\UrlBreaks}{\UrlAlphabet}
\g@addto@macro{\UrlBreaks}{\UrlDigits}
\makeatother

\usepackage{filecontents}
\begin{document}
\title {SvTPM: A Secure and Efficient vTPM in the Cloud}

%

\author{Juan Wang}
\affiliation{\institution{Wuhan University}}
\email{jwang@whu.edu.cn}

\author{Chengyang Fan }
\affiliation{\institution{Wuhan University}}
\email{cyfan@whu.edu.cn}

\author{Jie Wang}
\affiliation{\institution{Wuhan University}}
\email{iwangjye@gmail.com}

\author{Yueqiang Cheng}
\affiliation{\institution{Baidu USA Xlab}}
\email{strongerwill@gmail.com}

\author{Yinqian Zhang}
\affiliation{\institution{Ohio State University}}
\email{yinqian@cse.ohio-state.edu}

\author{Wenhui Zhang }
\affiliation{\institution{Pennsylvania State University} }
\email{wenhui@gwmail.gwu.edu}

\author{Peng Liu }
\affiliation{\institution{Pennsylvania State University}}
\email{pliu@ist.psu.edu}

\author{Hongxin Hu }
\affiliation{\institution{Clemson University}}
\email{hongxih@clemson.edu}

\pagestyle{plain}


\date{}

\begin{abstract}
Virtual Trusted Platform Modules (vTPMs) have been widely used in commercial cloud platforms (e.g. Google Cloud, VMware Cloud, and Microsoft Azure) to provide virtual root-of-trust for virtual machines. Unfortunately, current state-of-the-art vTPM implementations are suffering from confidential data leakage and high performance overhead.
In this paper, we present SvTPM, a secure and efficient software-based vTPM implementation based on hardware-rooted Trusted Execution Environment (TEE), providing a whole life cycle protection of vTPMs in the cloud.
SvTPM offers strong isolation protection, so that cloud tenants or even cloud administrators cannot get vTPM's private keys or any other sensitive data. In SvTPM, we identify and solve a couple of critical security challenges for vTPM protection with SGX, such as NVRAM replacement attack, rollback attacks, trust establishment, and a fine-grained trusted clock.
We implement a prototype of SvTPM on both QEMU and KVM.  Performance evaluation results show that SvTPM achieves orders of magnitude of performance gains comparing to the vTPMs protected with physical TPM. The launch time of SvTPM is 2600$\times$ faster than vTPMs built upon hardware TPM. In the micro-benchmarks evaluation, we find that the command execution latency of SvTPM is smaller than or equal to the existing schemes.
\end{abstract}

\maketitle

\section{Introduction}
A Trusted Platform Module (TPM)~\cite{tpmsummary,tpmlibrary,shao2015formal} is a physical chip providing a complete set of standard APIs for key management, integrity measurement, and trust chain management. Many existing software applications have been developed upon such an abstraction. However, in cloud computing platforms~\cite{armbrust2010a,lombardi2011secure}, tons of virtual machines may be running simultaneously on one physical machine. As a result, it becomes inefficient to use one single hardware TPM to provide a number of virtual machines with the aforementioned security services due to low I/O performance of TPM hardware, small Non-Volatile Random Access Memory (NVRAM) space, and a limited number of Platform Control Registers (PCRs).

To address this issue, Virtual Trusted Platform Modules (vTPMs) are proposed to provide virtual root-of-trust for virtual machines in commercial cloud platforms~\cite{liu2010cloud}.
Instead of using physical NVRAM, a vTPM uses a file, denoted as a {\em NVRAM file}, to hold secret keys.
Google~\cite{googlecloud}, VMware~\cite{vmware}  and Microsoft~\cite{hyper-v} have already adopted vTPMs in their cloud platforms. Similar to a physical TPM chip, a vTPM stores credentials for trust \& identity, crypto operations and integrity measurement. Accordingly, the security of vTPM itself is crucial to cloud  platforms and tenants~\cite{krautheim2010introducing, cucurull2014virtual}. Unfortunately, how to protect vTPM itself is still a challenging question.

 \vspace{4pt}
 \noindent \emph{\textbf{State-of-the-art solutions.}}
 Currently, there are three methods to protect a vTPM~\cite{berger2006vtpm, stumpf2008enhancing, whitaker2002denali,VirtinSpector2014}.  The first method is TPM-based protection~\cite{berger2006vtpm,stumpf2008enhancing} that uses a physical TPM to seal the NVRAM file of a vTPM to bind the vTPM to the physical TPM and provision static protection for TPM's secret data. However, this method cannot provide run-time protection for vTPM. Thus, malicious cloud administrators may obtain sensitive data of vTPM through shared memory. Meanwhile, it also incurs high I/O performance overhead due to the low-speed performance of the physical TPM. The second method~\cite{Matthew2012vTPM} uses a tailored lightweight VM to support virtual TPM functions. This method relies on VM isolation to provide security separation. However, this method has three major limitations: (1) It introduces high resources consumption because every virtual machine has a paired vTPM virtual machine. (2) In addition, cloud administrators may leak vTPM keys by accident or on purpose due to the weak isolation. (3) The I/O performance is restricted because it still needs to use the physical TPM as the root of trust. The third method is SMM-based protection~\cite{VirtinSpector2014}, which places the vTPM into CPU System Management Mode (SMM) to achieve strong isolation. This method has high performance overhead because entering the SMM mode requires suspending all other CPU cores.

\vspace{4pt}
\noindent \emph{\textbf{Key challenges.}} A promising alternative to existing solutions is to use Software Guard Extensions (SGX)~\cite{intelsgx} to achieve TPM-chip-free protection of vTPMs. The new vTPM solution with SGX protection will no longer be plagued by the restriction of a physical TPM's low speed I/O ports. However, since the kind of trust provided by SGX is very different from the kind of trust provided by a TPM chip, using SGX to protect a vTPM is significantly more challenging than only running the vTPM's functions  inside
an enclave (i.e. private regions of memory).

In particular, we observe that there are several critical security challenges that need to be addressed when we put a software-based TPM into an SGX enclave for strong isolation. To a large extent, these security challenges are introduced by the differences between the kind of trust provided by SGX and the kind of trust provided by a TPM chip. For example, although a TPM chip is tamper-proof, the NVRAM file of a vTPM -- when being protected by SGX -- is not really tamper-proof. The file can be replaced with another file by a compromised hypervisor or a malicious cloud administrator.

\begin{enumerate}
\item \textit{NVRAM Replacement Attacks}. The NVRAM file is used to store confidential data of a vTPM.  It is needed to ensure that the NVRAM file can only be accessed by a specific VM. However, attackers and malicious tenants may replace the NVRAM file of the vTPM to steal keys and data stored in the NVRAM file.
\item \textit{Trust Establishment between vTPM and SGX Platform}. Privacy CA (PCA) cannot rely on the Endorsement Key (EK) from TPM manufacturer to issue the Attestation Identity Key (AIK) certificate for the vTPM because the vTPM EK is generated by software instead of injecting into the TPM chip at manufacturing time. Hence, how to extend trust from SGX to vTPM so as to ensure the trust of TPM identity becomes a key issue.
\item \textit{Rollback Attacks of vTPM}. The rollback function of vTPM may be exploited by attackers to launch brute-force attacks. Thus, we need to design a new protection mechanism to prevent the rollback of sensitive values.
\item \textit{Fine-grained Trusted Clock}. Trusted time is associated with lockout persistence and authorization management of TPM. Although SGX Read Time-Stamp Counter (RDTSC) instruction is allowed in SGX2, its value can be manipulated by the privileged software and thus cannot be trusted by the enclave program. Moreover, providing a fine-grained secure clock based on other hardware modules~\cite{rajftpm} suffers from high performance overhead and weak security. Therefore, how to provide accurate and secure clock values through SGX is still a tough problem.
\end{enumerate}

\noindent \emph{\textbf{Our solution.}}  We present SvTPM, a secure and efficient software-based TPM, and propose the following methods to solve the above challenges in SvTPM:  (1) Designing a binding mechanism between vTPMs and VMs to prevent the NVRAM replace attacks. Therefore, the vTPM sensitive data and keys belonging to different VMs cannot be accessed by each other, and only the authenticated VM  can access their unique NVRAM file;  (2) Extending the trust from the SGX platform to vTPM and establishing strong links between vTPMs identity keys and SGX platform. It uses SGX Enhanced Privacy ID (EPID) to quote vTPM EK and AIKs, thereby passing the trust relationship from SGX to vTPM; (3) Provisioning two solutions: global software counter and SGX hardware counter, to defend against rollback attacks to vTPM; and (4) Presenting a fine-grained trust clock with Platform Service Enclave (PSE)  to satisfy with TPM 2.0 requirements.
Based on those approaches, SvTPM can provide security guarantees similar to that of a physical TPM chip for vTPMs in the cloud.

To demonstrate the effectiveness of our approach, we prototype SvTPM base on both QEMU and KVM platforms with real Skylake CPU and evaluate its performance. Evaluation results show that SvTPM has much better performance than using physical TPM chips while providing stronger isolation for vTPMs. Boot time for NVRAM using SvTPM is 2600 times faster than using hardware TPM. Execution of commands is also comparable with hardware TPM. Compared with vTPM without SGX protection, SvTPM incurs a small, barely noticeable performance degradation. Furthermore, SvTPM is fully backward compatible, so that cloud tenants do not need to modify their applications and services when they use virtual TPMs.

SvTPM is a practical solution that addresses the above key challenges while using SGX to protect vTPMs and then implements secure software-based vTPMs for the cloud.

\vspace{4pt}
\noindent \emph{\textbf{Our contributions.}} Our main contributions are as follows:
\begin{itemize}
\item We identify security challenges for vTPM protection with SGX, such as NVRAM replacement attacks, trust establishment between vTPM and SGX Platform, rollback attacks to vTPM, and fine-grained trusted clock.

\item We propose SvTPM, a secure and efficient software-based TPM, to provide vTPM run-time protection and strong isolation based on SGX. In SvTPM, we design a binding mechanism between vTPM and VM to prevent the NVRAM replace attacks. In addition, we introduce an approach to establishing trust between SGX and vTPM. Furthermore, we present two rollback attack defense solutions and a fine-grained trusted clock mechanism.

\item We implement a prototype of SvTPM based on both QEMU and KVM, and evaluate its performance. The prototype comprises 43,000 newly developed lines of code.
Experimental results show that SvTPM has a better performance than using a physical TPM while providing strong isolation and secure protection for vTPMs in the cloud.

\end{itemize}

\section{Background}
\label{sec:background}
In this section, we introduce the background of Trusted Platform Module, virtual Trusted Platform Module and SGX technologies used in this work.

\subsection{Trusted Platform Module and TPM 2.0}

Trusted Platform Module (TPM)~\cite{tpmsummary,han2018a} is a hardware chip which servers as trusted computing base. Users can use it to measure the integrity of their system and keep sensitive files confidential. Nowadays, TPMs are provided in many commodity servers, desktops, and laptops.

NVRAM file stores persistent state associated with TPM. Persistent data includes root keys such as Endorsement Key (EK), Attestation Identity Key (AIK), Storage Root Key (SRK),  state information about a machine, measured values of a system and user's keys, which are generated based on the root keys. NVRAM file is controlled by owners and can be configured to control read and write capabilities separately. This means that only authorized users can use or manage NVRAM files. However, storage space for NVRAM files is limited so that it can not store a large amount of data.

TPM provides a set of PCRs to store measured values. When a measurement is extended to a PCR, the value is hashed together with the previously stored values in the PCR. Then the PCR is updated with the hashed results. A small change will affect all subsequent extension values. Also, specific PCR values can be reproduced only when the same values are extended in the same order. Thus updating the values in this way makes it easy to find integrity violations in the current trust chain. In Trusted Computing Group (TCG) specification, there are 16 PCRs~\cite{bade2010scalable}. And users can measure key modules and store measured values to these PCRs.

TPM often uses Low Pin Count (LPC) bus to transmit data. Though some TPM 2.0 chips support I$^2$C and PCI, these protocols are still hard to meet standards of high-speed concurrent encryption and decryption operations.

Currently, TCG has released TPM 2.0 specification~\cite{tpmmain, tpmlibrary}, which follows ISO standard (i.e. ISO/IEC 11889:2015). Compared with TPM 1.2, TPM 2.0 has many advantages. (1) TPM 1.2 only has one hierarchy in storage. TPM 2.0 has three persistent hierarchies. It has hierarchy in platform, storage, and endorsement. Each of these hierarchies owns at least one root of trust, such as Endorsement Primary Seed (EPS), Storage Primary Seed (SPS) and Platform Primary Seed (PPS).  Additionally, TPM 1.2 only supports SHA-1 and RSA, while TPM 2.0 supports all kinds of cryptography algorithms. TPM 2.0 includes an algorithm identifier that allows TPM to use any algorithm design without changing its specifications. Hence, TPM 2.0 can be easily integrated with various cryptography algorithms, such as Chinese Commercial Cryptography algorithms, SM2, SM3, and SM4. TPM owns and generates different types of keys. Among these keys, EK is an identity key of TPM. In TPM 1.2, there is only one EK, which is built within the chip and cannot be modified. In TPM 2.0 specification, it has multiple keys (i.e. EK, AIK, SK, Signing Key (SIGK), Binding Key (BK), Authentication Key (AK) and Legacy Key (LK)). EK is initially generated by TPM chip manufacturer. It can be regenerated by the platform manufacturer and platform owner. In other words, original EK could be updated to newer EK versions.

In addition, TPM 2.0 unifies the way in which entities in TPM are authorized. Besides traditional password and HMAC authentication methods, authentication method based on policy authorization is added. It allows multiple policies to be used. This enhances the security of the keys. Moreover, TPM 2.0 also enhances its robustness. In the TPM 2.0 specification, important data can be sealed to a PCR value approved by a particular signer.

\subsection{vTPM}
vTPM ~\cite{berger2006vtpm} is a virtual TPM mainly used in cloud platform which serves as root-of-trust for virtual machines.
vTPM service in Xen is designed as a set of secure, stand-alone domains managed by the hypervisor. Each of these domains has a mini-OS and their dedicated functions. Each domain ( i.e. virtual guest) has a software-simulated TPM running in the vTPM domain called DOMU. vTPM manager runs in a privileged domain called DOM0, where DOM0 is responsible for the creation and management of the vTPM instances and establishing interaction between vTPMs and a physical TPM.

The vTPM based on QEMU-KVM is implemented with two modes, TPM passthrough mode, and full vTPM mode. In TPM passthrough mode, hardware TPM is exposed to a virtual machine in the form of a virtual TPM instance. All operations of this virtual TPM instance are passed to the underlying TPM hardware. And the hardware TPM does all computing and storage work. Full vTPM mode provides a virtual machine with a vTPM implementation that is completely detached from the physical TPM. In this approach, TPM functions are implemented by designing software TPM backend and the externally linked Libtpms.

Our work focuses on the implementation of full vTPMs mode in QEMU-KVM architecture. In this architecture, KVM works as virtualization infrastructure for the Linux kernel, which turns Linux kernel into a hypervisor. QEMU is responsible for device emulation, which emulates hardware devices for VMs. VM sends TPM requests to QEMU, which hands them over to vTPM devices emulated by QEMU. Libtpms implements functions of TPM as a comprehensive software library. vTPM leverages Libtmp library for its functionality implementation. vTPM calls APIs of libtpms. Libtpms processes the requests, then returns the results to QEMU, which is then passed to VM via TPM driver. An NVRAM file stores important state information of the vTPM, including keys and PCR values. The advantage of this architecture is that it supports multiple virtual machine instances and easy migration functionality of these virtual machine instances. However, the security of vTPM cannot be guaranteed due to lacking hardware isolated environment.

\subsection{Software Guard Extensions}

SGX~\cite{costan2016intel, sgxguide, sgxreference} is a set of instructions and mechanisms for memory accesses added to Intel CPU processors. It provides enclaves to ensure confidentiality and integrity of key applications.

{\bfseries Enclave.} Enclave is an SGX based trusted execution environment for applications. Code and data in the enclave reside in a protected physical memory area called enclave page cache (EPC). Accesses to the code and data in enclaves are protected by SGX access control mechanism. When loaded into DRAM, data in the EPC page is protected by the granularity of one cache line. On-chip memory encryption engine encrypts and decrypts cache lines written to and extracted from the DRAM. Also, the integrity of enclave memory is guaranteed, which means that memory modifications and rollbacks can be detected. The non-enclave code cannot access enclave memory, while enclave code could access external untrusted DRAM.

{\bfseries Enclave life cycle.} An enclave is created when ECREATE instruction is processed. In this process, a free EPC page is allocated as SGX enclave control structure (SECS). Then the EADD instruction associates this allocated page to the enclave. SGX records which pages are associated with which enclave by their virtual address and permissions of the pages. Then SGX performs security checkings for page access permission checking and page allocation checking. After success allocation of enclave pages, EINIT instruction generates secure hashes for remote attestation. After instances in enclaves are finished, EREMOVE instruction is executed to de-allocate the EPC pages.

{\bfseries Performance overhead.} Performance overhead is introduced when code is executed in enclaves in three scenarios as follows. (1) Since privileged instructions cannot be executed within the enclave, threads must exit the enclave before the system call. The enclave transition needs to perform a series of check and update, including TLB refreshes, which incurs performance overhead. (2) Because the Memory Encryption Engine (MEE) must encrypt and decrypt the cache line, overhead for writing memory and caches misses is introduced. (3) Page swapping between EPC and unprotected DRAM is expensive because the confidentiality and integrity of the evicted EPC pages and the freshness of the pages brought back to EPC should be guaranteed. In order to avoid address translation, it is necessary to interrupt all enclave threads and refresh the TLB. These interruptions introduce performance overheads.

\section{Motivations and Security Challenges}
 In the cloud, functions supported by SGX abstraction library is not comparable to that of TPM module even though SGX provides attestation and several similar features as TPM.  This is mainly due to the following reasons:

\begin{enumerate}
\item
SGX supports the protection for specific applications instead of their execution environment. It utilizes enclaves for isolation between applications, rather than between applications and the execution environment they are executing on.

\item  Memory of EPC page provided by SGX is limited, which is restricted to 128M~\cite{brenner2016securekeeper}. When available EPC pages are insufficient, least recently used EPC pages are swapped to available ones. During the swapping stage, EPC pages are transferred into untrusted DRAM pages. These pages' virtual addresses are beyond address range of Processor Reserved Memory (PRM). During this transferring process, memory encryption and decryption and TLB flush are involved. Thus, very high performance overhead incurs.

\item SGX has limited support for security services and user-friendly APIs, such as key management, crypto operations, random number generation operations, and secure storage functions. Thus, the diversity of cryptography algorithms becomes a straggler for sharing security services~\cite{tian2019a}.

\end{enumerate}

In SvTPM, our goal is to build a system, which supports secure vTPMs with high performance using SGX. Our SvTPM provides a secure virtual TPM for cloud users. With the vTPM, users can verify the trust of key modules in virtual machines, reuse existing services (e.g. BitLocker), and implement a variety of personalized security services. However, we have identified many security challenges that need to be solved to implement a secure and efficient virtual TPM with SGX in cloud.

\subsection{NVRAM Replacement Attacks}

vTPM does not have the physical NVRAM of TPM chips so that each vTPM needs a unique protected file to store its sensitive data. The file in the host representing the flash of the hardware TPM is used to store the vTPM's persistent information, such as seeds, keys and PCRs, which is called NVRAM. Since NVRAM stores the confidential data  of vTPM, its security is crucial to cloud platform. However, software systems are naturally different from hardware systems, it is necessary to consider any potential security issues caused by software implementations in SvTPM. We have identified the following security requirements that
NVRAM needs to meet in SvTPM to achieve a similar security
strengths provided by the hardware TPM:

\begin{enumerate}
\item  Data stored in the NVRAM files must be encrypted and securely stored;
\item NVRAM files cannot be accessed  by the other enclaves except
their owners;
\item The NVRAM file of a vTPM can only be accessed by an authorized VM.
\end{enumerate}

vTPM servers as root-of-trust of VM instances. SvTPM needs to provide a unique vTPM for each VM of a cloud. In the cloud, it must ensure that the NVRAM can only be accessed by the specific VM, thus preventing confidentiality leaks. However, an attacker can replace the NVRAM file and legally launch the vTPM so as to steal the user's credentials. Hence, how to protect the NVRAM and prevent NVRAM replace attacks is a key challenge for SvTPM.

\subsection{Trust Establishment between vTPM and SGX platform}

Another key problem for SvTPM is the establishment of the trust between vTPM and SGX Platform. TPM performs some authentication operations using the attestation identity key (AIK). TPM needs to interact with the PCA to obtain an AIK certificate before using the AIK. The privacy CA will issue AIK certificates based on EK certificates issued by the hardware manufacturer to TPM. Since the EK certificate is generated by TPM manufacturer, this certificate can provide identification for TPM. Different from hardware TPM, SvTPM is implemented by software. Thus, the following two questions need to be answered in SvTPM:

\begin{itemize}

\item Where to obtain the EK certificate of vTPM?
\item What are requirements for issuing the EK certificate?

\end{itemize}

In a software-based vTPM, it lacks trusted EK and AIK. The EK and AIK are generated by the software. Hence how to rely on SGX platform to establish a trust relationship between vTPM and SGX so as to ensure the trusted identity of the vTPM is another challenge.  \\

\subsection{Rollback Attacks to vTPM}

The rollback operation is a basic function provided by the cloud platforms for users. A user can take a snapshot of a VM in advance and then restore the VM to the previous state after an exception occurs. For vTPM architecture, the data and states of vTPMs also need to be rollbacked when VMs are rollbacked. Otherwise, it will cause that the state of vTPM and VM is inconsistent. This feature provides a great convenience for users, but it also means that the VM environment of a cloud platform is different from the physical machine~\cite{dai2017rollsec, xia2012defending, matetic2017rote}. For the SvTPM, adversaries may use the rollback function to threaten the password security of the SvTPM.

\begin{figure}[t]
\begin{center}
\includegraphics[width=3in]{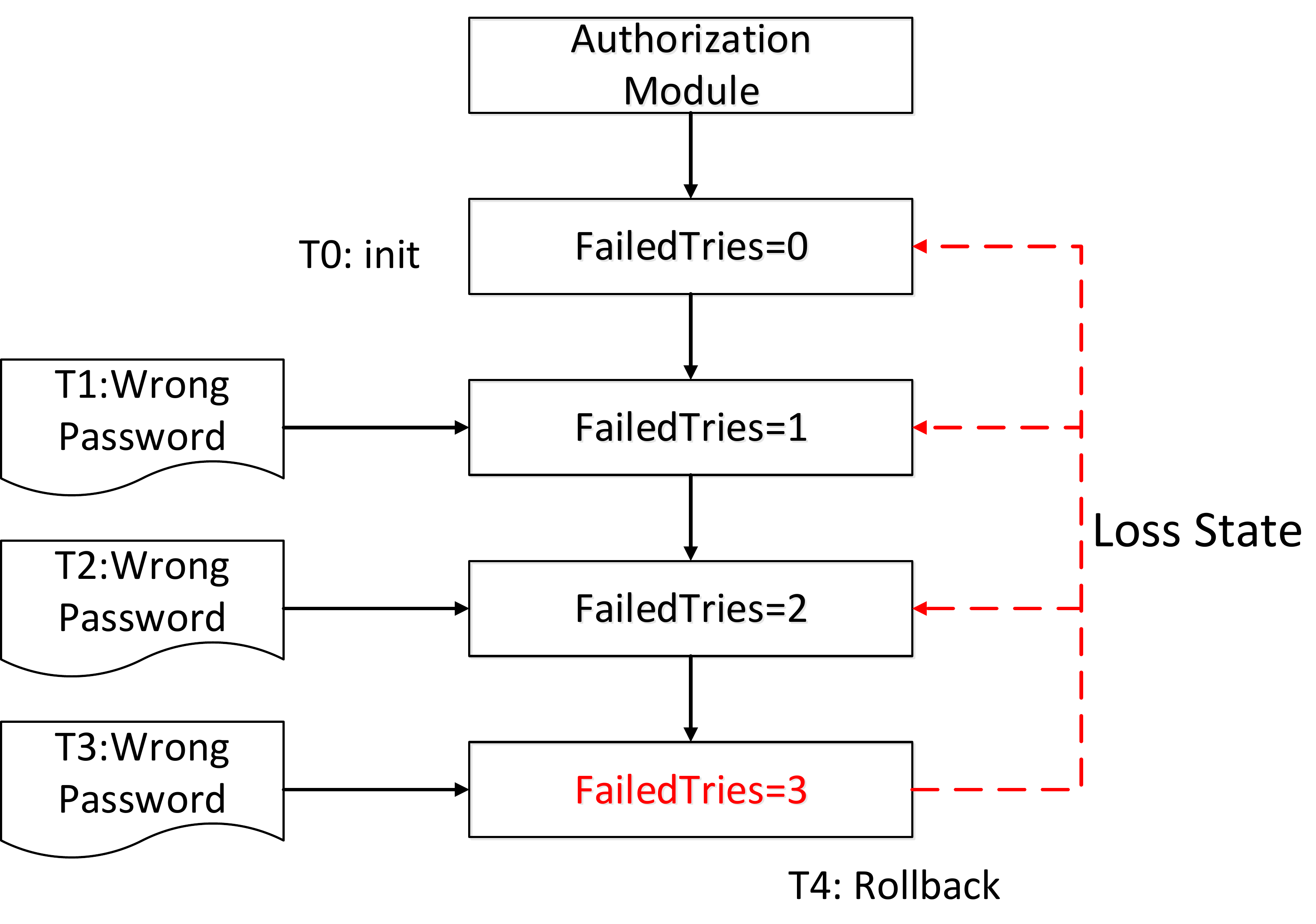}
\end{center}
\caption{Process of Rollback Attack.}
\label{fig:Rollback}
\end{figure}

FailedTries in TPM, for example, is used to prevent dictionary attacks against TPM. Many of the functions in the TPM require a user to authorize and, if the key entry is incorrect, the value of failedTries will increase depicted in Figure~\ref{fig:Rollback}. When a threshold value is reached, the TPM enters the locked state. In a cloud platform, if malicious users use the rollback function to rollback SvTPM after SvTPM triggers the lock,  they can bypass the protection strategy and successfully launch a brute-force password attack on SvTPM. Hence, it is a challenge to design a new protection mechanism to prevent the rollback of sensitive values in vTPM.\\

\subsection{Fine-grained Trusted Clock}

There are two reasons why TPM needs a trusted clock. First, TPM needs to measure the lockout duration for some specific attacks. For example, TPM is required to implement a lockout mechanism to protect against so-called "dictionary attacks", where an attacker tries numerous passwords until one succeeds. If passwords are entered incorrectly more than $n$ times during the authentication process, the TPM enters the lockout state and refuses any service for a pre-determined period of time.
Second, the trusted clock can be used for time-stamped authorizations, such as creating a valid authentication for a pre-defined period of time. For example, the TPM can create a key valid for a day, which will become unusable when the timer expires.

TPM 2.0 specification requires a clock with millisecond granularity. Then, a TPM can use such a clock to measure intervals of time for lockouts and time-bound authorizations.

Unlike TPM chips, vTPM lacks the hardware clock. We want to provision the hardware clock based on SGX, but SGX does not also have a hardware clock. SGX RDTSC (i.e. read time-stamp counter) instruction provides secure clocking service, which is not permitted in the enclave mode in SGX1. Although this instruction is allowed in SGX2, its values can be manipulated by the privileged software and thus cannot be trusted by the enclave program. Moreover, providing a secure clock based on other auxiliary hardware modules~\cite{rajftpm},  is limited to the specific architecture and low performance. Therefore, how to provide a fine-grained trusted clock through SGX becomes a challenge.

\section{Our Approach}
In this section, we describe our threat model, design goals and the architecture of SvTPM.
\subsection{Threat Model}
We consider a powerful adversary who has superuser access to systems and physical hardware. Software, including OS, SeaBIOS, and
other system software are not trusted. However, considering the management of virtual machines is controlled by hypervisors in the cloud, we assume hypervisors are benign-but-vulnerable, while they may be subjected to bugs.

We only trust CPU and the code running in an enclave. We assume that the adversary is unable to subvert cryptographic primitives correctly implemented and is unable to subvert the security guarantees of SGX on any machines. However, we assume that attackers have the abilities
to do the following things:

\begin{itemize}
\item Analyze source code to determine sensitive data structures,
and then directly analyze NVRAM that stores vTPM nonvolatile
data to obtain sensitive data;

\item with root privileges, use dynamic debugging and tracking
program execution process to find sensitive data during VM
running;
\item Change and replace the NVRAM file path associated with a
VM at startup to access NVRAM information belonging to
other vTPM;
\item Take a snapshot and rollback the states of VMs and vTPMs, and conduct rollback attacks.
\end{itemize}

Side-channel attacks~\cite{Chen2017,fu2017sgx-lapd,vanbulck2018foreshadow} on SGX are out of scope of our threat model. However, a number of solutions have been recently proposed to mitigate such attacks~\cite{chen2018racing,gruss2017strong,shih2017t-sgx,shinde2016preventing}.

\subsection{Design Goals}

To address the above security challenges, we propose SvTPM, which is an SGX-based vTPM system. The design goals of SvTPM are articulated as follows:

\begin{itemize}
\item Minimize the changes of the existing vTPM architecture so that cloud tenants do not need to modify their applications and services when they use virtual TPMs;
\item Protect the confidentiality and integrity of the vTPM in run-time from information leaking and data tampering;
\item Ensure the secure storage of sensitive files of vTPM (i.e. NVRAM files), and prevent files from being accessed illegally;
\item Establish a one-to-one binding relationship between NVRAM files and VMs to mitigate NVRAM replacement attacks;
\item Prevent credentials of vTPM from being leaked when hypervisors are comprised;
\item Minimize performance overhead and achieve a higher access speed than physical TPMs.
\end{itemize}

\subsection{High-Level Architecture of SvTPM}

\begin{figure}[t]
\begin{center}
\includegraphics[width=3in]{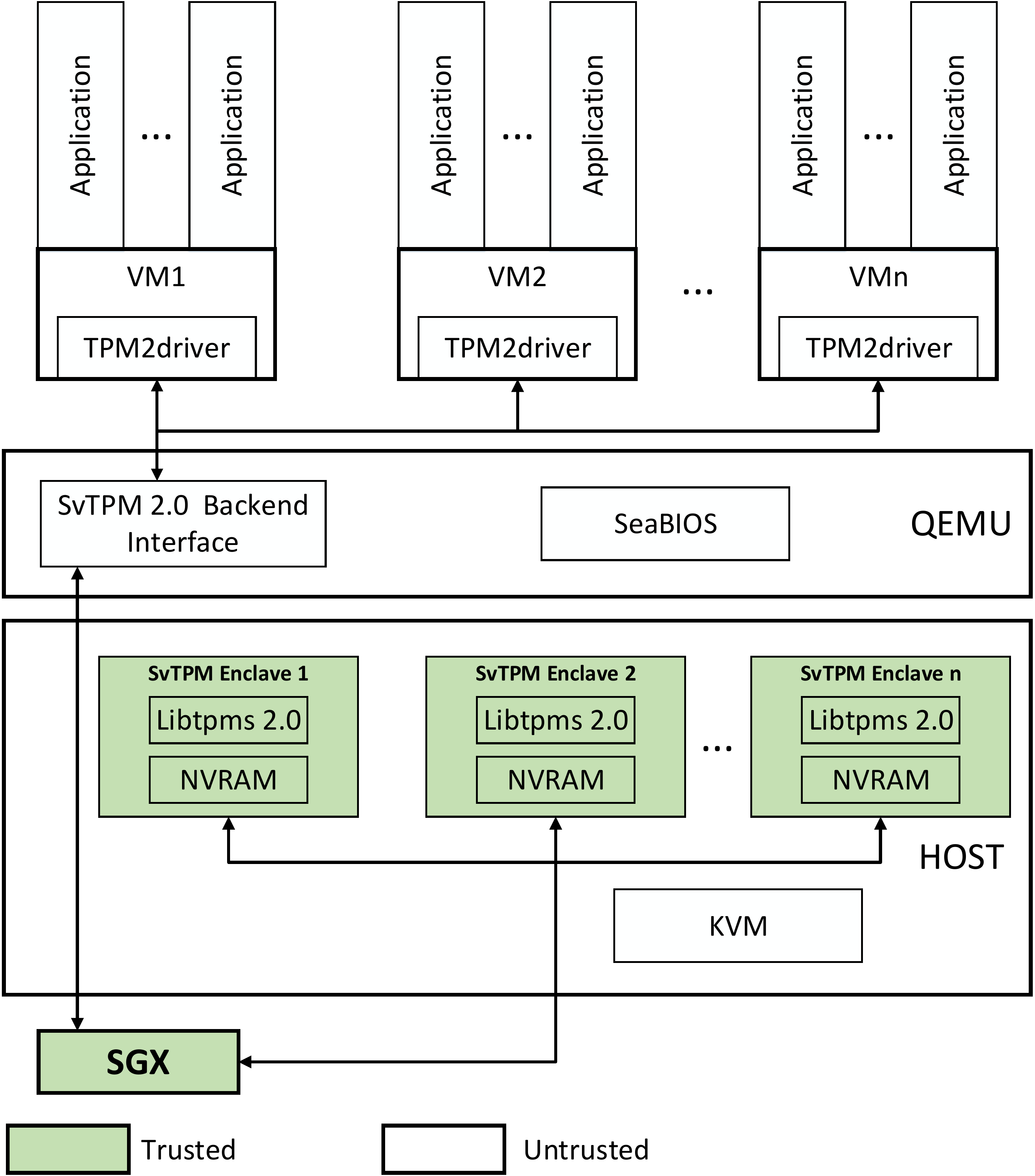}
\end{center}
\caption{High-Level Architecture of SvTPM. Four main components are highlighted: (1) SvTPM enclave instances; (2) SvTPM 2.0 backend interface; (3) SeaBIOS; and (4) Tpm2driver.}
\vspace{-0.1cm}
\label{fig:arch}
\end{figure}

\begin{figure}[t]
\begin{center}
\includegraphics[width=3in]{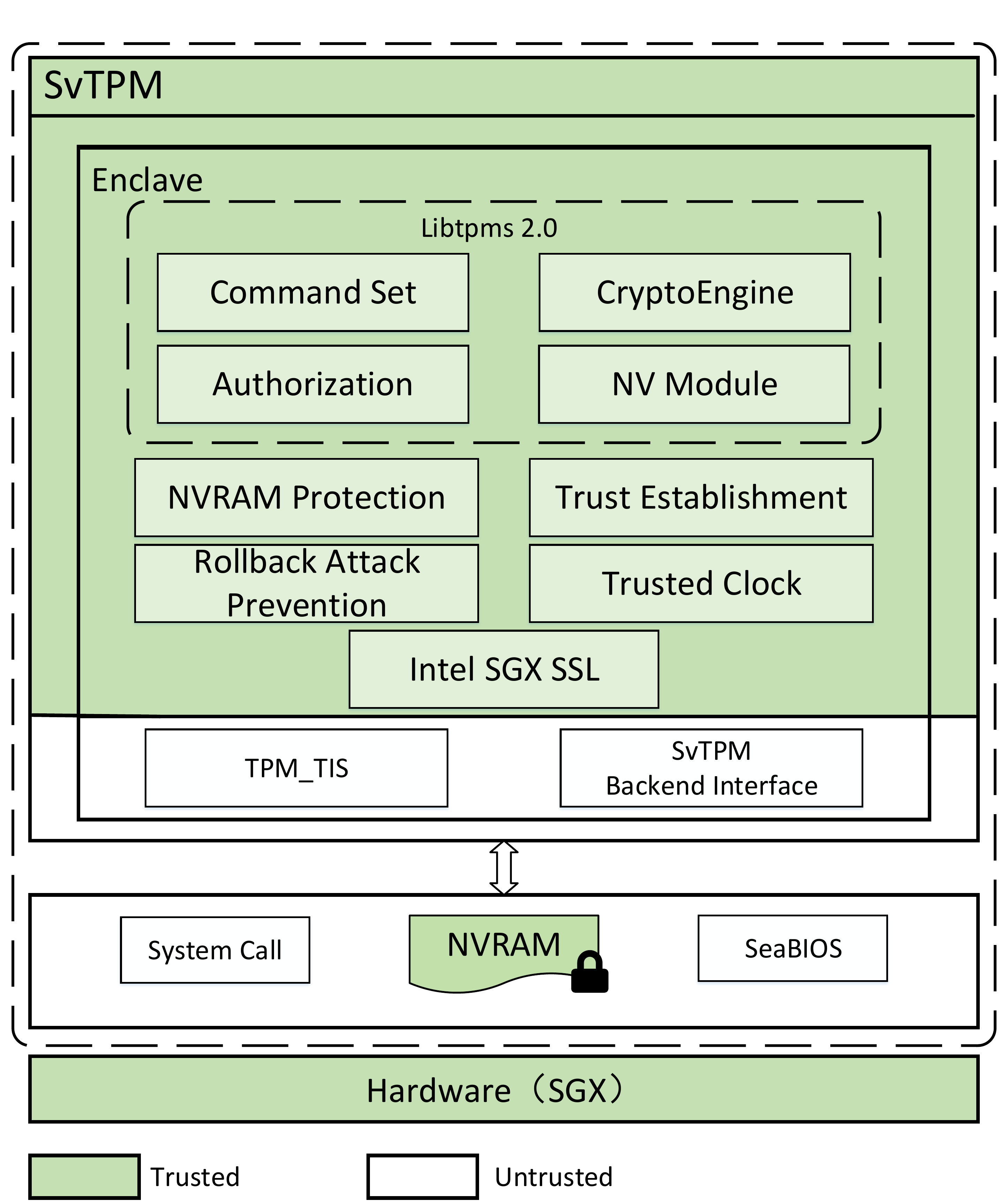}
\end{center}
\caption{Trusted Modules in SvTPM Enclave. }
\vspace{-0.3cm}
\label{fig:svtpm}
\end{figure}

A high-level architecture of SvTPM is shown in Figure 2. It includes the following major components: SvTPM enclave instances, SvTPM 2.0 backend interface, SeaBIOS, and Tpm2driver. The SvTPM enclave includes the Libtpms 2.0 library and the NVRAM of a vTPM, which are sealed and isolated in an SGX enclave. The SvTPM 2.0 backend interface is responsible for creating SvTPM enclave instance, providing command processing interface, etc. SeaBIOS is a virtual BIOS serving for guest OSes. The TPM2driver module provides the frontend interfaces to access virtual TPMs. In SvTPM, each virtual machine (VM) is provided with a unique SvTPM enclave instance to implement a secure vTPM.

When a VM get started, QEMU is responsible for initializing its devices. At this stage, an SvTPM device is initialized. Meanwhile, SvTPM generates an enclave for the vTPM and makes sure the vTPM initialization instruction is safely executed within the enclave. The enclave has the same lifetime as the VM is only destroyed if the VM is shut down or restarted. When a user executes a TPM command with TSS (TPM Software Stack) in the VM, it is passed to QEMU, which is responsible for passing the command to the enclave, where specific instructions are safely executed and results are returned by the enclave to QEMU, which then returns to the upper driver.

Due to the limited memory size of SGX EPC, we need put the key modules of the vTPM to the SvTPM enclave that is divided into two logical parts: the trusted part and the untrusted part depicted in Figure~\ref{fig:svtpm}. The trusted part is used to perform confidential operations (such as key generation, crypto operations, etc.), to store sensitive data, and to solve security challenges. Its code and data are protected in the SGX enclave. In SvTPM system, the trusted parts include the Libtpms library, NVRAM files, Intel SGX SSL, and four security modules, which are responsible for confidential operations. The untrusted parts help with non-confidential operations. Code and data of untrusted parts are not protected by SGX. For SvTPM, the untrusted parts include TPM\_tis, vTPM backend interface, system\_call, and SeaBIOS. The trusted parts and the untrusted parts cannot directly interact with each other. In order to securely communicate between the trusted parts and the untrusted parts, we need to define the interface file between them, which is called the EDL file. The EDL file defines OCALL function where the untrusted parts call the trusted parts and ECALL function where the trusted parts call the untrusted parts.

Libtpms 2.0, the shared library, is the key trusted module of SvTPM. It can provide all the features and command sets for TPM 2.0.
Libtpms 2.0 consists of a command set module, a crypto engine module, an authorization module, and an NV module. The command set module implements various commands defined in the TPM 2.0 specification. The authorization module implements Enhanced Authorization (EA) of TPM 2.0, which manages all objects or entities of the TPM (including the TPM's hierarchies) and the authorization policies associated with them. The NV module implements NVRAM file operation, such as creating, resetting, reading, and writing, and other NV operations. The crypto engine module packages the implementation of the encryption and decryption algorithms provided by the TPM and implements them through the calling interface provided by OpenSSL. However, due to the enclave mechanism, OpenSSL cannot be used directly, but Intel provides the Intel SGX SSL library for use.

Since the Libtpms 2.0 modules support core functions of the system, its code and data need to be protected. Whenever QEMU creates a VM, it first creates an enclave through SGX and loads Libtpms 2.0 into the enclave for isolation. After loading Libtpms 2.0, enclave measures hash of Libtpms 2.0 to verify its integrity. If its integrity is not compromised, the Libtpms 2.0 code runs in the EPC. With SGX's physical memory isolation protection and memory access control mechanism, it can ensure that other software, including privileged software, cannot access enclave data. Meanwhile, data in EPC pages are encrypted by the memory encryption engine. This data can only be obtained through hardware memory attacks. Therefore, this mechanism protects the vTPM memory code and data at run-time.

The NVRAM file is the other trusted component of SvTPM, which stores keys, PCR values, seeds, and other private data. When a vTPM device is created, the NVRAM file will also be created. Since the important data of the vTPM is stored in the NVRAM file, its secure storage is crucial for the security of the vTPM. Whenever the vTPM performs a command operation, SvTPM will update the corresponding data to the NVRAM file to save the state of the vTPM and then uses SGX sealing operation to seal it to disk. When data is needed, the NVRAM file sealed on the disk will be first unsealed and then loaded into the EPC for usage.

We add four security modules in the SvTPM enclave, which are NVRAM protection module, trust establishment module, rollback protection module, and trust clock module. With these security components, we address the security threats mentioned above, such as NVRAM replacement attacks, trust establishment between vTPM and SGX platform, rollback attacks to vTPM, and trusted clock. The following sections provide in-depth details of these four security modules.

\subsection{NVRAM Protection}

In SvTPM, we use the seal mechanism of SGX to encrypt the data stored in NVRAM files that can also ensure that the NVRAM files cannot be accessed by the other enclaves except their owners. SGX's seal mechanism utilizes AES-GCM to encrypt data and provides two key derivation policies, KEYPOLICY\_MRENNCLAVE and KEYPOLICY\_MRSIGNER. KEYPOLICY\_MRENCLAVE and KEYPOLICY\_MRSIGNER use the enclave measurement and signer measurement registers to derive the keys, respectively. In SvTPM, we choose to use the KEYPOLICY\_MRSIGNER key derivation policy to meet the requirements of NVRAM secure storage and NVRAM isolation. The enclaves for vTPMs have different seal keys.

For the NVRAM replacement attack, the fundamental cause is the lack of strong binding between VM and NVRAM. As mentioned above, data stored in NVRAM in the SvTPM architecture is the result of seal operation. After the seal operation, all the data stored in NVRAM is encrypted. If a user's enclave can be signed with his or her secret key during the enclave generation phase, the KEYPOLICY\_MRSIGNER key derivation policy makes each NVRAM sealed by a separate key. This key derivation policy satisfies the requirements of secure storage and isolation, but it is still not sufficient to establish a binding relationship between the VMs and NVRAMs. An attacker who can replace the NVRAM enclave file can still achieve the NVRAM replacement attack if the corresponding enclave file is replaced at the same time.

Aiming at defending NVRAM replacement attacks, establishing a binding relationship between VM and NVRAM file is necessary. If a VM can access another VM's NVRAM file, many of the persistence data in NVRAM belonging to one VM can be replaced by another.
The software applications that rely on NVRAM persistence data are prone to such attacks. In a physical machine, the binding relationship between the physical machine and the TPM chip is determined by the physical connection. This connection relationship enables the TPM chip to be capable of being independent of the physical machine and can provide proof of identity to the physical machine. In the software implementation, we need additional mechanisms to implement the binding relationships due to the absence of this physical binding relationship.

In order to solve the problem of binding VM and vTPM. We design a binding scheme for signing VM images and enclave based on the user's private key. When a user creates a virtual machine with a virtual TPM, we ask the user to generate her/his own unique key pair and then use that key pair to sign the user-specific enclave file. After this step, we can ensure that a separate key is obtained after using the KEYPOLICY\_MRSIGNER key derivation strategy to ensure the isolation between NVRAMs of different users. In addition, we will use this key to sign the image file of the VM and check the integrity of the VM image when the VM is powered on. Through the above two protection strategies, we can establish a binding relationship between the VM image and the NVRAM file. In addition to establishing the binding relationship between VM and NVRAM, we also design a binding verification scheme based on a remote third-party. When the vTPM starts up, after the NVRAM is properly unsealed, its identity information is passed to the cloud platform through the trusted channel. When the VM is started, the vTPM measures the VM and sends the measurement of the VM to the cloud platform after the measurement process ends. When the cloud receives two pieces of information, it compares the enclave's measurements with the VM's measurements and determines whether the binding relationship is correct.

\subsection{Trust Establishment}

\begin{figure}[t]
\begin{center}
\includegraphics[width=3in]{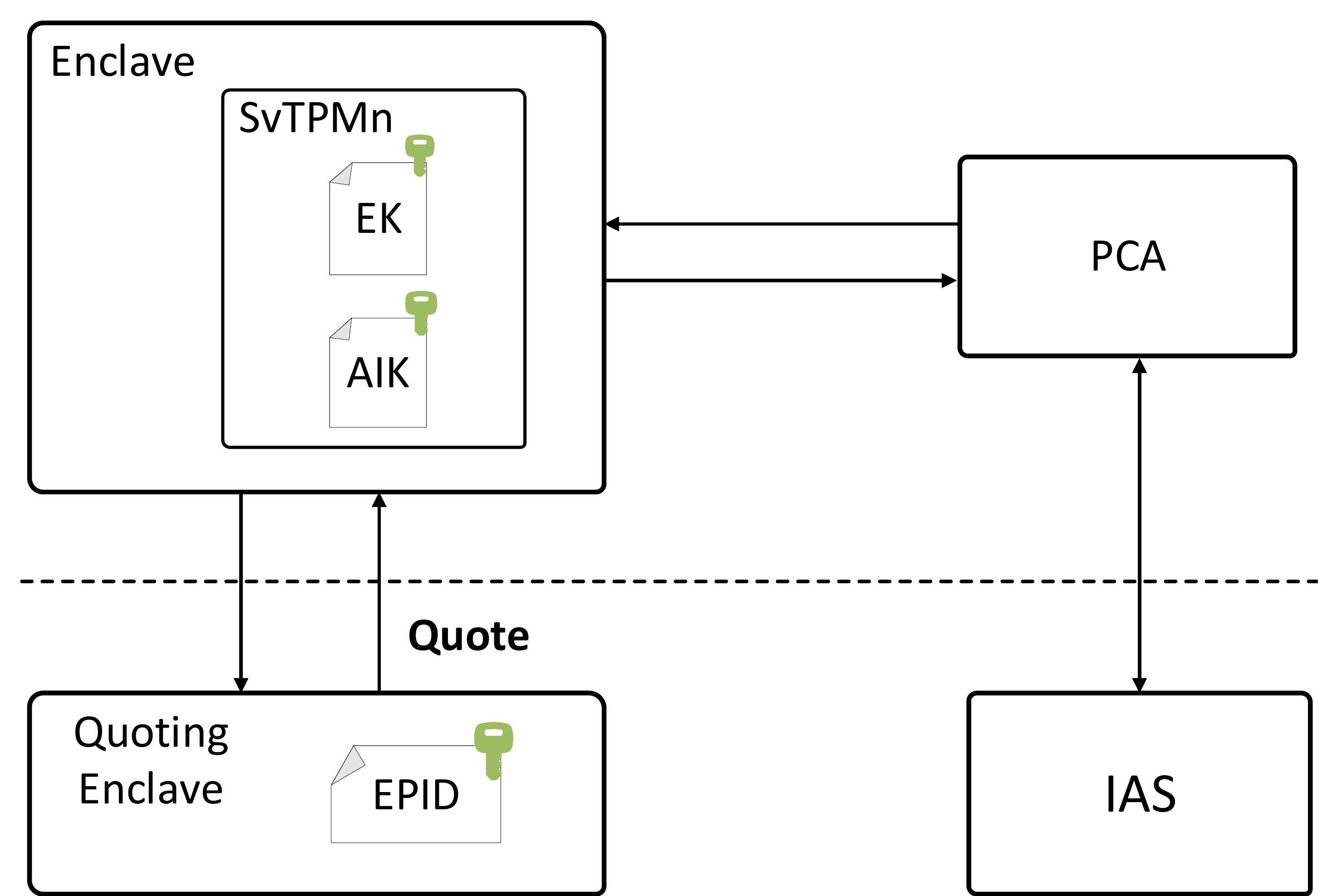}
\end{center}
\caption{Trust Establishment.}
\vspace{-0.3cm}
\label{fig:trustest}
\end{figure}

In the vTPM architecture based on hardware TPM, AIK from hardware TPM and TPM quote operation can be used to bind the EK certificate of vTPM to the underlying hardware platform. The quote operation of TPM signs the EK of vTPM with the current state of PCRs and the hash of EK. PCRs hold state information of software and hardware of the platform, hence this operation is equivalent to bind EK of vTPM with the underlying platform. Since we no longer use hardware TPM in SvTPM, we need to investigate SGX-based alternatives for SvTPM.

Let's first consider the security features provided by SGX itself. SGX provides integrity and confidentiality for enclaves~\cite{epid}. If we put a vTPM in an enclave, it means the vTPM is measured for integrity on startup and cannot be influenced by malicious interference at run-time. This shows that we have narrowed the vTPM attack to the vTPM code and SGX mechanism itself. As long as we can ensure the security of both, we can trust the vTPM. After the above analysis, we define the following two conditions for issuing EK certificates: (1) vTPM runs on SGX; and (2) Enclave measurement value of vTPM is invariant.

To satisfy those two conditions, we provide a trust establishment solution depicted in Figure~\ref{fig:trustest}. When a PCA ( Privacy CA ) needs to issue an AIK certificate for a vTPM, it, as a challenger, first remotely attests the vTPM. In this scenario, SGX attestation key will generate a signature based on the Enclave's MRENCLAVE and the hash value of vTPM EK. Then, the enclave of vTPM requests Quoting enclave for SGX signature. The Quoting enclave generates EPID to sign the vTPM values. The signature values from the Quoting enclave is then sent to the PCA. PCA verifies the signature by Intel Attestation Service (IAS). When the verification succeeds, it provides us two pieces of information: the vTPM is running on hardware SGX; and the integrity of the enclave where the vTPM is located is not comprised. As long as both of these conditions are satisfied, the vTPM can be proved to be within the security boundary, and the privacy CA can issue a certificate for it.

\subsection{Rollback Attacks Prevention}

To mitigate the rollback attack on vTPMs, we propose the following two defense mechanisms based on software and hardware, respectively:

\begin{figure}[t]
\begin{center}
\includegraphics[width=3in]{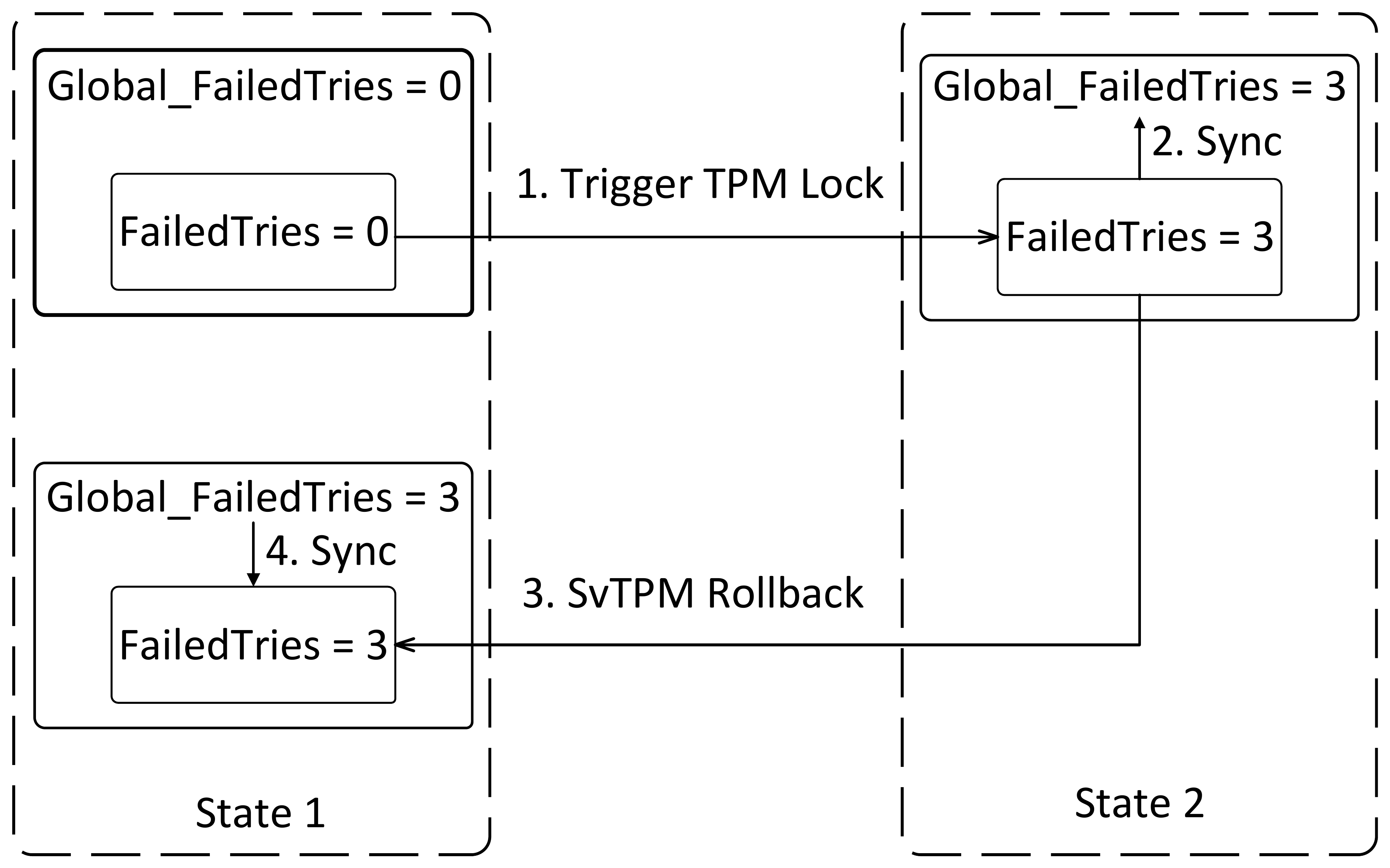}
\end{center}
\caption{Rollback Protection Synchronization Mechanism Based on Software.}
\label{fig:rollbacksw}
\end{figure}

\begin{figure}[t]
\begin{center}
\includegraphics[width=3in]{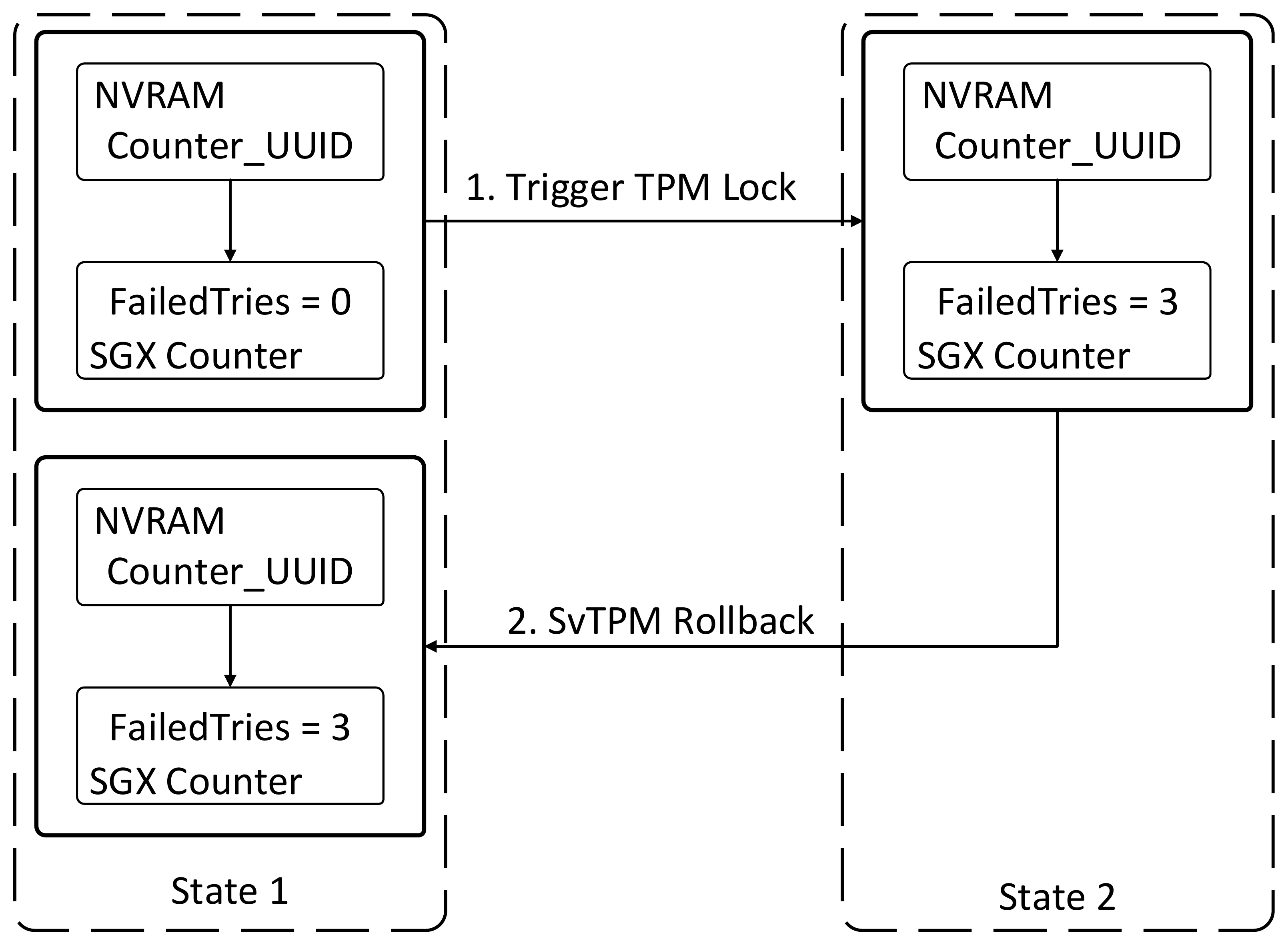}
\end{center}
\caption{Rollback Protection Synchronization Mechanism Based on SGX Monotonic Counter.}
\vspace{-0.3cm}
\label{fig:rollbackhw}
\end{figure}

\begin{itemize}
\item \textbf{Software-based mechanism}. As shown in Figure~\ref{fig:rollbacksw}, we design a synchronization mechanism to synchronize the non-rollback data in SvTPM. With this scheme, the rollback attack against vTPM can be defeated by ensuring that the rollback operation cannot affect the non-rollback data in SvTPM. As an example of the FailedTries mentioned above, we add a Global\_Failedtries value outside the rollback space of vTPM status, which is not affected by the rollback operation. This value is synchronized using the FailedTries value in the normal operation of vTPM. However, when the rollback operation occurs, because this value is not affected by the rollback operation, FailedTries can be synchronized using this value at the end of the rollback operation to ensure that FailedTries in vTPM space are not affected by the rollback operation.

\item  \textbf{Hardware-based mechanism}. As is shown in Figure~\ref{fig:rollbackhw}, we use hardware monotonic counters provided by Intel ME to represent non-rollback data like FailedTries in vTPM. When vTPM is first initialized, we use the \textit{sgx\_create\_monotonic \_counter()} function provided by SGX SDK (i.e. Intel ME) to apply the monotonic counter, which returns a Counter\_UUID. We need to save this value in NVRAM for further manipulation. When a vTPM needs to operate on FailedTries, we use the previously saved Counter\_UUID to access and modify the hardware counter. SGX provides three access strategies for the monotonic counter: the first one has the same signature key; the second is an enclave that has the same measurement value; and the third is an enclave that has the same signature key and the measurement value. This access policy ensures that the hardware monotonic counters used by SvTPM are not affected by the enclaves created by an external malicious party. In addition, due to the limited number of hardware counters, 256 in total, we provide an additional interface for users to destroy SGX hardware counters to prevent the abuse of SGX hardware counters.
\end{itemize}

Both of the above solutions can solve the rollback attack problem, and both have advantages and disadvantages. The advantage of the software-based solution is that it can prevent the rollback attack at less cost. Since our software program is implemented primarily through a couple of synchronous operations, it only adds a few synchronized statements to the original and barely increases the performance overhead. The disadvantage of the software-based solution is that if the synchronization process is maliciously interrupted, the protection strategy can be bypassed. If the synchronization process is interrupted before Golbal\_FailedTries synchronizes FailedTries, FailedTries will incorrectly stay in the state of the previous snapshot, causing it to be rolled back incorrectly. The advantage of the hardware-based solution is that it does not have the problem of data being rolled back due to an interruption in the middle of the rollback in the software scheme. Because the essence of the hardware-based solution is to isolate the non-rollback data from the NVRAM and put it into the SGX monotonic counter, it cannot be affected by external software operations. Even if it is interrupted in the middle of the rollback operation, its data cannot be rolled back, because the SGX monotonic counter will not be affected.

Meanwhile, there are three main disadvantages of the hardware-based solution: (1) There is a significant performance penalty associated with the operation of a hardware SGX counter; (2) In the TPM specification, FailedTries values are reduced after a certain recovery time, but we cannot simulate this situation after using hardware monotonic counters to represent the FailedTries values because SGX does not provide an interface for diminishing monotonic counters. In this case, we must modify the relevant semantics in the TPM specification. When FailedTries return to 0, the SGX counter is destroyed and reapplied; and (3) Reapplying the SGX counter can inevitably lead to the change of Counter\_UUID. If the snapshot is taken before the change of Counter\_UUID and recovered after that change,  the SGX counter cannot be accessed.

Both solutions for addressing the rollback attack problem have trade-offs, and users can choose a suitable scheme according to their own usage scenarios. For example, if users have high-performance requirements, they can choose the software-based mechanism; if users are not sensitive to performance and do not have additional protection policies to ensure the continuity of rollback operations, they can choose the hardware-based mechanism.

\subsection{Fine-grained Trusted Clock}
The construction of SvTPM requires a trusted clock, which unfortunately cannot be supported directly by SGX enclaves. In this section, we present a solution to address such a problem.

 SGX can obtain trusted clock functions through the SGX Platform Service~\cite{sgxpse}. The Platform Service Enclave (PSE) can provide a timer with the Protected Real-Time Clock (PRTC) of Converged Security and Management Engine (CSME) to application enclaves. This trusted time service can be used by an application enclave running on an offline platform to track the amount of time passed since a previous reading of the timer. It also returns a nonce, which is trusted according to Intel standards as long as the nonce does not change. However, this trusted clock service provides clock values in seconds relative to the reference point. This is inconsistent with the millisecond granularity requirement of the TPM 2.0 specification.

The SGX RDTSC instruction provides nanosecond-granular timestamp counters. However, the SGX RDTSC instruction is not permitted in the enclave mode of SGX1. Although this instruction is allowed in the enclave mode of SGX2, its values can be manipulated by the privileged software and thus cannot be trusted by the enclave program. Thus, we cannot use the SGX RDTSC instruction as a secure clock in SvTPM.

In our approach, we combine the coarse-grained PRTC with a fine-grained software clock. The software clock is implemented by running a separate software thread inside the same enclave, which executes in a loop and increments a counter at a constant rate.  We adopt the method proposed by Chen et. al.~\cite{Chen2017} to realize such a software clock. Therefore, in SvTPM, when a coarse-grained clock is required, we only leverage the trusted clock values through the PSE. When a fine-grained clock is needed, we launch a software thread that generates a fine-grained clock inside the enclave and periodically checks the PRTC to correct its values.

\section{Evaluation}
In this section, we evaluate our approach with the following major goals:
\begin{itemize}
\item Evaluating the performance of SvTPM when it is launched. We tested the NVRAM launch time (Figure 7).
\item Evaluating and analyzing the performance of SvTPM when creating the RSA key. The results are shown in Figure~8.
\item Evaluating the performance of SvTPM in some TPM common commands, such as seal, unseal, sign, signature verification, encrypt, decrypt, PCR read, and PCR extend. These operations basically cover the common operations of the TPM~(Figures~9,~10, and 11).
\item Demonstrating the performance of SvTPM in practical applications. As shown in Figure 12, we tested the performance of SvTPM in BitLocker encryption and decryption.
\item Evaluating the resource consumption of SvTPM. We compared the memory usage of vTPM in XEN and SvTPM (Figure~13).
\end{itemize}

\subsection{Experiment Environment}

We implement SvTPM, and in this section we describe how we evaluate its performance and discuss evaluation results.
Our experimental environment was set up on an E3 server with Intel Xen e3-1280 v6 CPU and 16GB of memory. The model of hardware TPM in our experimental environment is NATIONZ TPM 2.0. Our host operating system is Ubuntu 18.04, SGX SDK version is 2.3, QEMU version is 2.6.0, and the SeaBIOS version is 1.10.0. The operating system version of the VM is ubuntu 14.04, and its memory is 2GB. The number of new lines of code added for developing SvTPM is over 43,000.

\begin{figure}[h!tb]
\begin{center}
\includegraphics[width=3in]{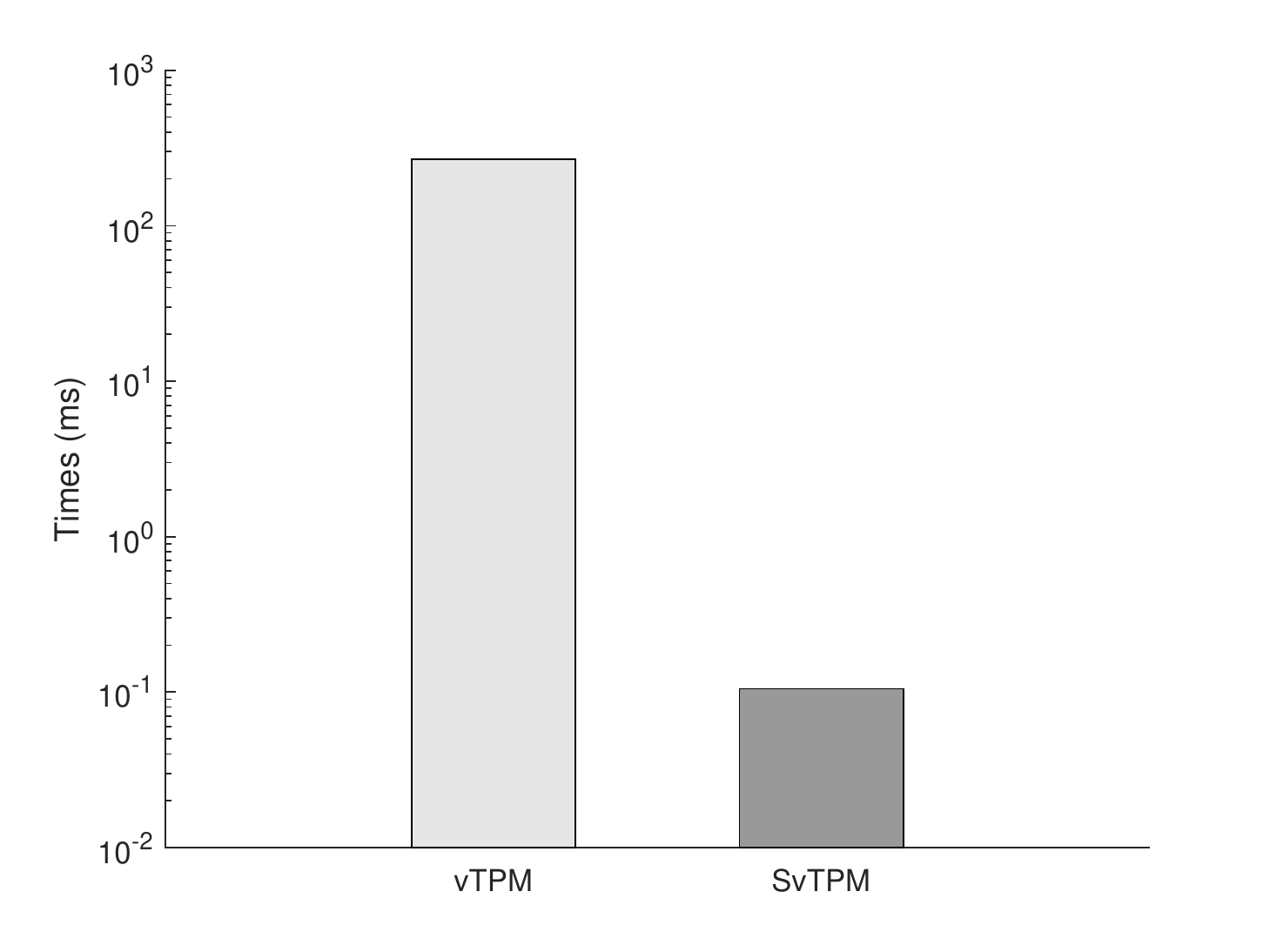}
\end{center}
\vspace{-0.1cm}
\caption{NVRAM Launch Time. The NVRAM launch time of SvTPM is about 2600+ times faster than that of vTPMs built upon  hardware TPM.}
\vspace{-0.3cm}
\label{fig:launch-time}
\end{figure}

\subsection{NVRAM Launch Time}

In this vTPM architecture based on hardware TPM, the first load of NVRAM needs to decrypt the encrypted key of NVRAM with TPM device in the host machine, and then decrypt NVRAM with the key. In the SGX-based SvTPM architecture, NVRAM encryption and decryption are replaced by SGX's seal and unseal operations. We have tested the NVRAM launch time in the above two different architectures for 100 times. Experimental results are shown in Figure~\ref{fig:launch-time}.  The NVRAM launch time spent by the SGX-based SvTPM architecture is about 0.015ms, while the vTPM architecture based on hardware TPM took 267ms to launch NVRAM file.  In contrast, the NVRAM launch time of SvTPM is about 2600+ times faster than that of vTPMs built upon hardware TPM.

The reason is that the vTPM architecture based on hardware TPM has two performance limitations compared to the SGX-based SvTPM architecture. First, the vTPM architecture based on hardware TPM needs to decrypt the key first and then decrypt NVRAM file with the key in two steps, while in SvTPM, only one step is needed to decrypt NVRAM file.  Second, the step of password decryption needs to interact with the hardware TPM. Since the hardware TPM is a low-speed device, performance loss in this stage is significantly large.

\subsection{Performance of TPM Commands}
In order to test the performance of SvTPM, several commands commonly used in TPM were selected and tested on vTPM, SvTPM and hardware TPM respectively. Here, vTPM refers to vTPM architecture without SGX protection. In order to obtain more accurate experimental results, we use an average of 100 times of executions for performance evaluation.

\begin{figure}[h!tb]
\begin{center}
\includegraphics[width=3in]{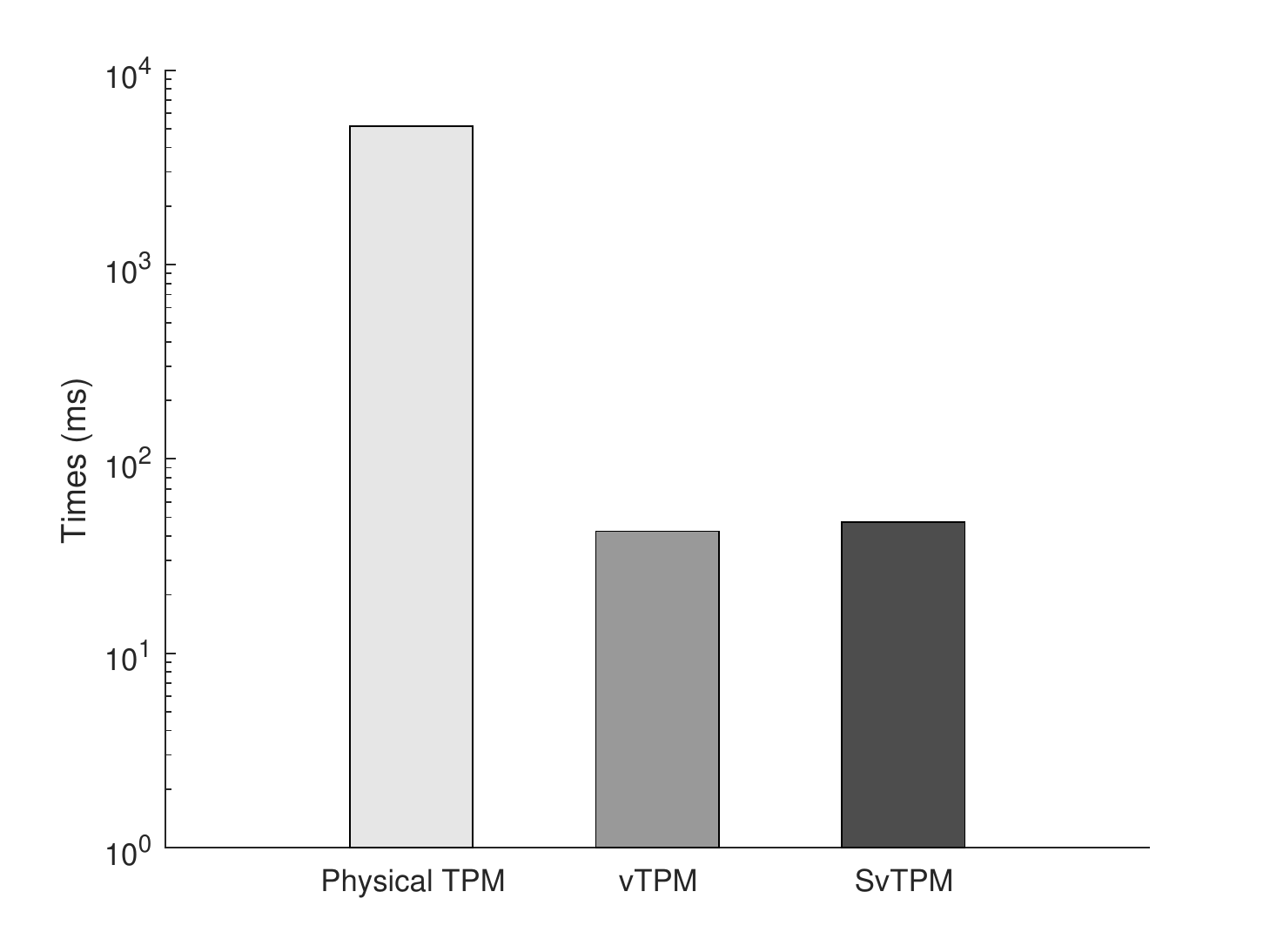}
\end{center}
\vspace{-0.1cm}
\caption{Performance of Creating RSA Keys. The SvTPM speed of creating RSA keys is about 110 times faster than that of the hardware TPM. }
\vspace{-0.3cm}
\label{fig:RSA}
\end{figure}

Figure~\ref{fig:RSA} shows the cost of creating an RSA key in three environmental settings. Creating an RSA key in TPM consists of finding prime numbers, creating public and private parts of the key, and encrypting the private part with the root key. Creating an RSA key is a relatively time-consuming operation in physical TPM, which took an average of 5.17 seconds in the tested physical TPM, compared to 42.44 ms in vTPM and 47.19 ms in SvTPM. We can see that time consumption of this command can be greatly reduced in the vTPM. According to our analysis, the advantages of the vTPM environment are mainly in two aspects. First, the speed of generating random numbers is faster than that of hardware TPM. Second, the process of verifying prime numbers in the virtual environment can take advantage of the computing power of CPU to bring significant performance improvement. In addition, we can see that the time consumption of SvTPM is about 4.7ms more than that of vTPM without SGX protection, which brings about 11\% additional performance overhead.

\begin{figure}[h!tb]
\begin{center}
\includegraphics[width=3in]{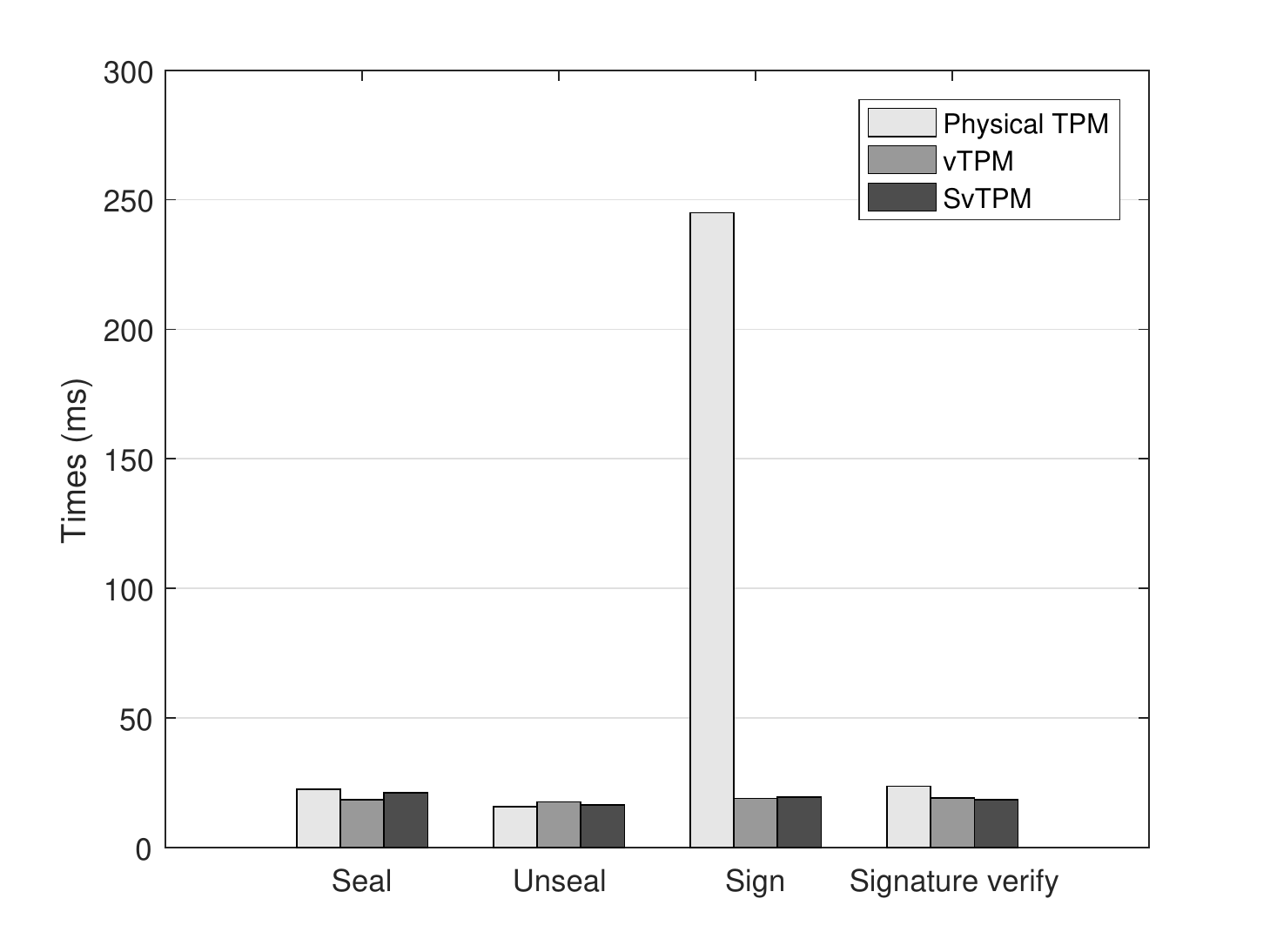}
\end{center}
\vspace{-0.1cm}
\caption{Performance of Common Commands. The signature speed of SvTPM is about 13 times faster than hardware TPM. }
\label{fig:common}
\vspace{-0.3cm}
\end{figure}

Figure~\ref{fig:common} shows the results of testing other four main operations including \textit{seal}, \textit{unseal}, \textit{sign}, and \textit{verify}. It is obvious that vTPM and SvTPM can significantly reduce the time loss in the \textit{sign} operation. The signature speed of SvTPM is about 13 times faster than that of the hardware TPM. The benefits of the virtual environment are not apparent in the other three operations, even though the \textit{unseal} operation consumes a little more time than that in physical TPM. In addition, we can see that the performance of vTPM and SvTPM in these four commands are competitive. This indicates that SvTPM only brings little overhead to these operations.

\begin{figure}[h!tb]
\begin{center}
\includegraphics[width=3in]{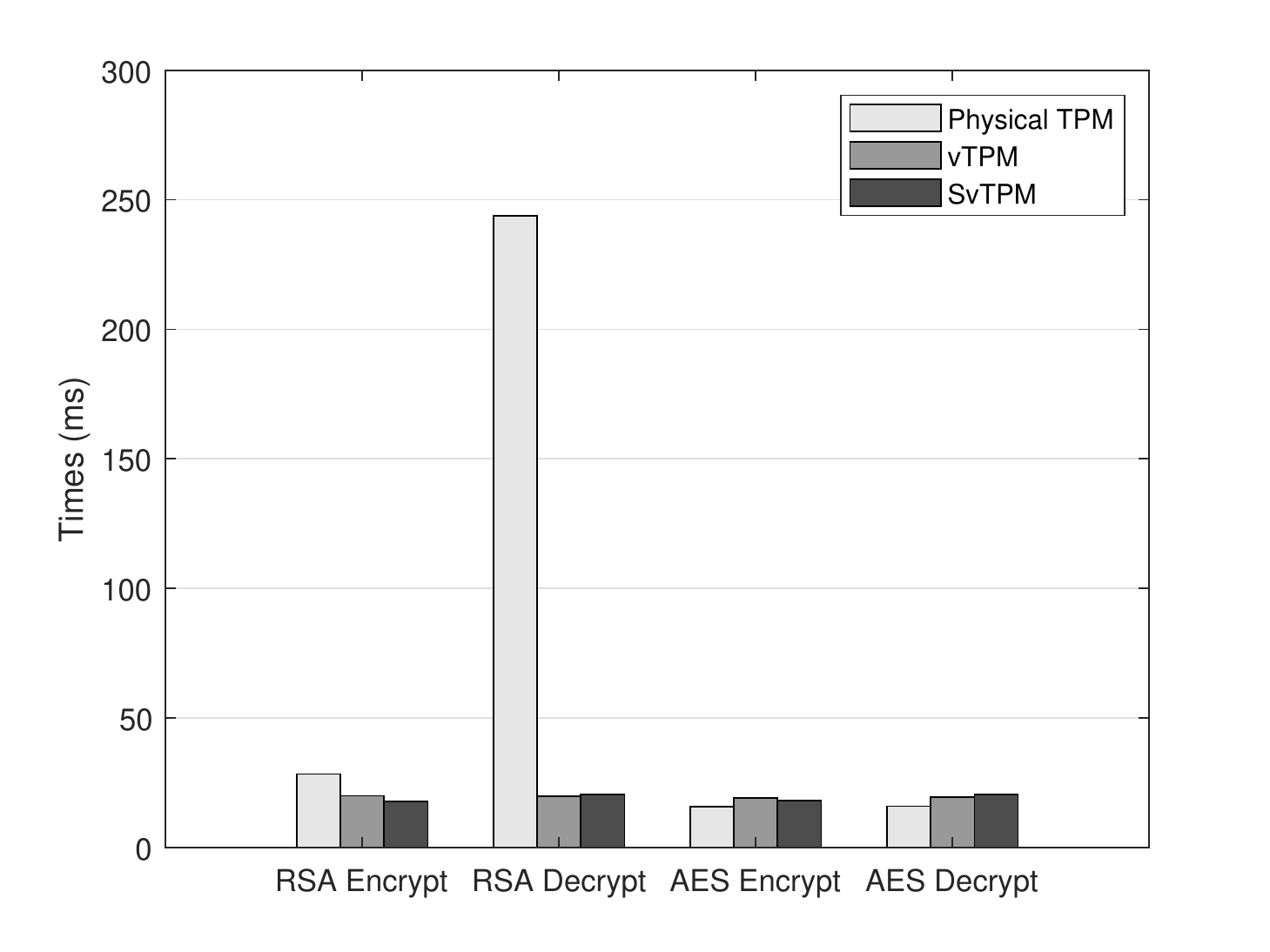}
\end{center}
\vspace{-0.1cm}
\caption{Performance of Encryption and Decryption. The decryption speed of SvTPM is about 12 times faster than the hardware TPM. }
\vspace{-0.3cm}
\label{fig:Encryption}
\end{figure}

Figure~\ref{fig:Encryption} depicts results of testing encryption and decryption operations of both AES and RSA. The results are similar to the results shown in Figure~\ref{fig:common}. SvTPM and vTPM offer significant performance improvements in the RSA decryption, which is about 12 times faster than that of the hardware TPM. However, the AES encryption and decryption in SvTPM and vTPM perform a little worse than those in hardware TPM. Finally, in terms of RSA encryption, SvTPM and vTPM can provide about 35\% performance improvement.

\begin{figure}[h!tb]
\begin{center}
\includegraphics[width=3in]{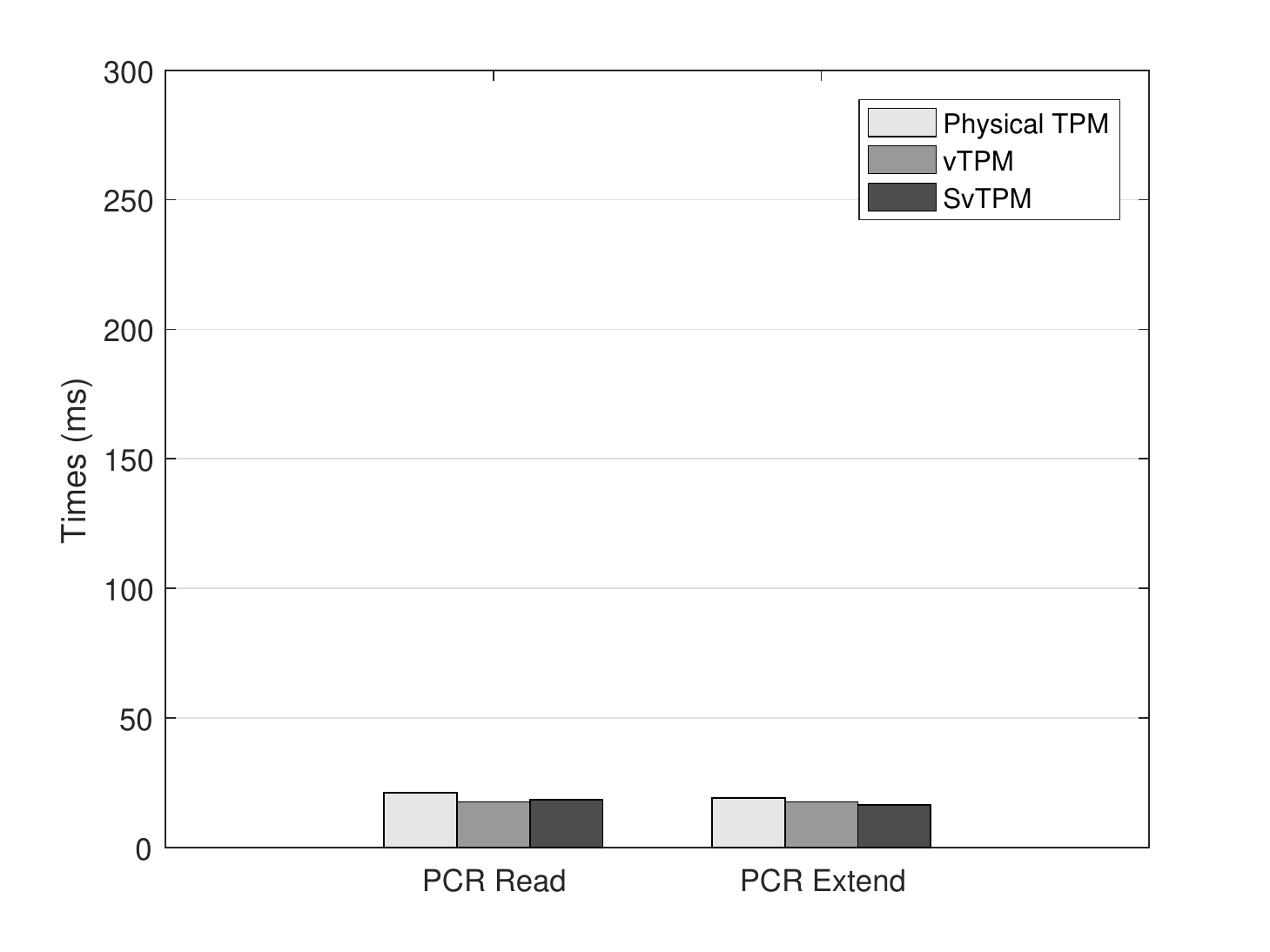}
\end{center}
\vspace{-0.1cm}
\caption{Performance of PCR Read and PCR Extend. }
\vspace{-0.3cm}
\label{fig:PCR}
\end{figure}

Figure~\ref{fig:PCR} depicts the test results of three different architectures in PCR-related operations. From the figure, we can see that hardware TPM spends a little more time on PCR read and PCR extend, and vTPM and SvTPM have similar performance in these two operations.

\begin{figure}[h!tb]
\begin{center}
\includegraphics[width=3in]{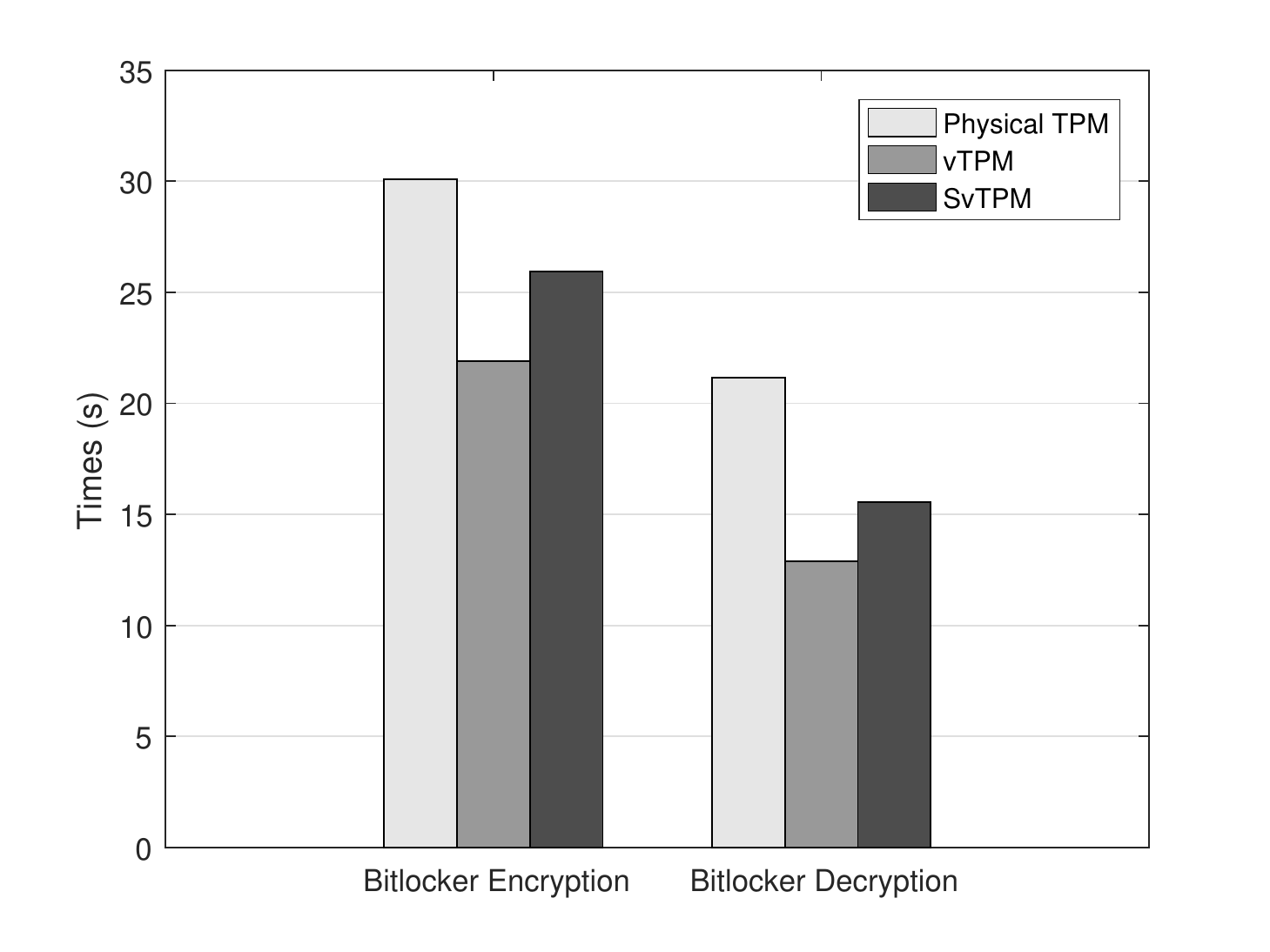}
\vspace{-0.1cm}
\end{center}
\caption{Performance of BitLocker encryption and decryption. }
\vspace{-0.3cm}
\label{fig:Bitlocker}
\end{figure}

In addition to testing the basic TPM commands, we also tested the impact of different architectures on real applications. As shown in Figure~\ref{fig:Bitlocker}, we tested the performance of BitLocker in Windows. In our test environment, we allocated an additional 2GB hard drive for Windows and tested the performance of BitLocker encryption and decryption for the drive in three different architectures. The evaluation results are shown in Figure~\ref{fig:Bitlocker}. As can be seen from the figure, BitLocker takes longer to do encryption than do decryption on the whole. In addition, among the three architectures, physical TPM takes the most time, vTPM takes the least time, and the performance of SvTPM is in the middle. The reason is that the keys for disk encryption and decryption are protected by TPM SRK.

From the above analysis, we can draw a conclusion that the command execution time of SvTPM is equal to or better than the existing schemes. Hence, it is worth considering SvTPM, which also provides run-time security protection for vTPMs in cloud.

\subsection{Memory Usage}

\begin{figure}[h!tb]
\begin{center}
\includegraphics[width=3in]{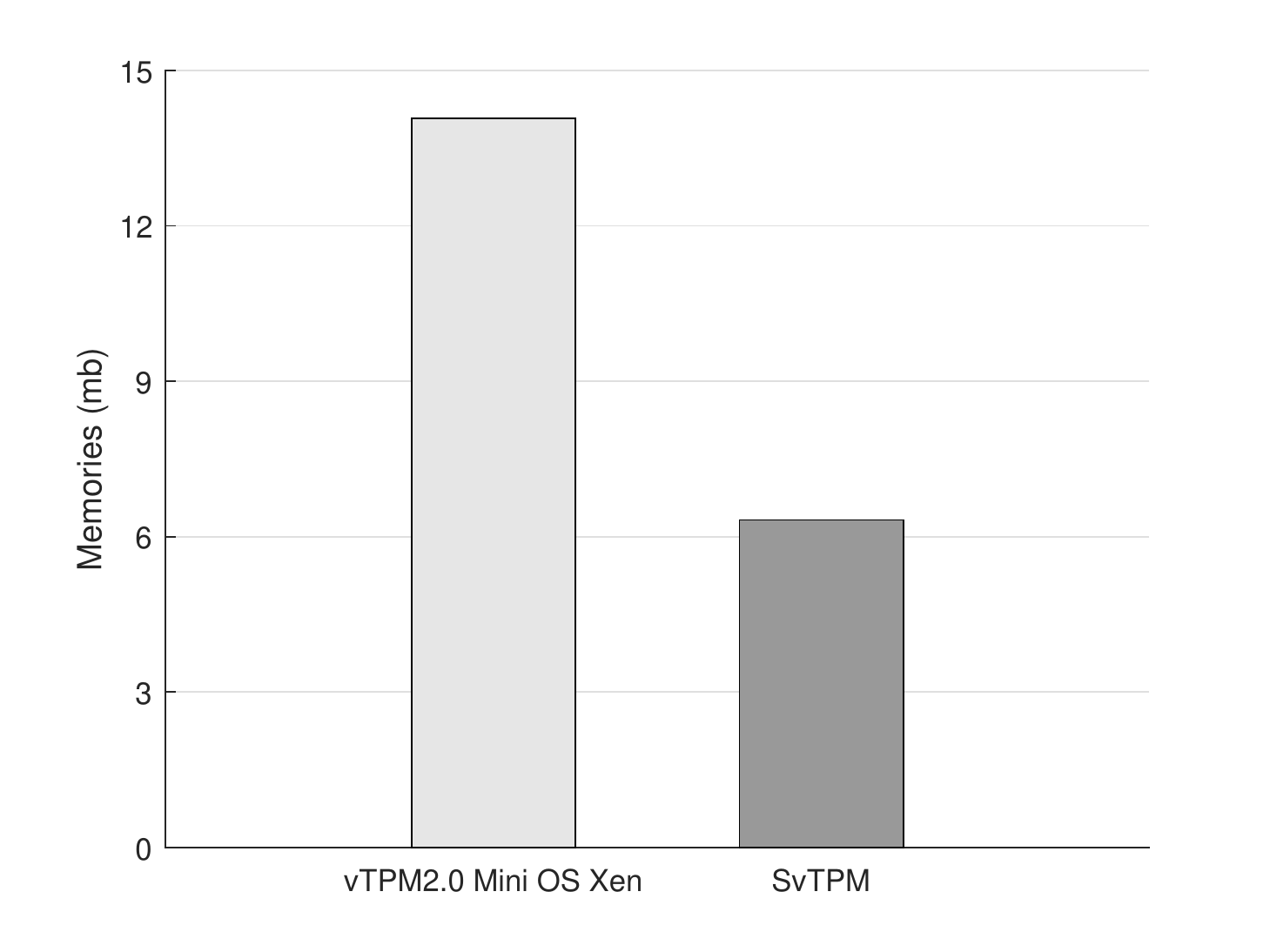}
\vspace{-0.1cm}
\end{center}
\caption{Memory Usage. The memory consumption of SvTPM is less than half of that of vTPM 2.0 MiniOS Xen.}
\vspace{-0.3cm}
\label{fig:Memory}
\end{figure}

In terms of memory consumption, we evaluate vTPM architecture in Xen. The vTPM architecture in Xen requires a MiniOS to run vTPM instances, which result in the need to occupy both MiniOS and vTPM memory. The experimental results are shown in Figure~\ref{fig:Memory}, which depicts that the memory consumption of SvTPM  is less than half of the vTPM architecture in Xen.

\begin{table*}
\caption{Comparison Between SvTPM and Other Similar Works}
\centering
\begin{threeparttable}[b]
\resizebox{\textwidth}{!}{
\begin{tabular}{cccccccc}
\toprule
\multirow{1}{*}{Solution} & \multirow{1}{*}{Trust Anchor} & \multirow{1}{*}{Supported Architecture} & \multirow{1}{*}{Secure Storage} & \multirow{1}{*}{NVRAM Replacement Protection} & \multirow{1}{*}{Trust Establishment} & \multirow{1}{*}{Rollback Attacks Protection} & \multirow{1}{*}{Fine-grained Trusted Clock}   \\
  \midrule
  TPM~\cite{tpmsummary} & TPM & Physical PC  & Physical NVRAM& N/A & TPM & N/A & \checkmark \\
  fTPM~\cite{rajftpm} & TrustZone & ARM & eMMC RPMB & N/A & N/A & \checkmark & $\times$ \\
  eTPM ~\cite{sunetpm} & TPM+SGX & Xen & Sealing & N/A  & TPM $\rightarrow$ SGX $\rightarrow$ eTPM & $\times$ & $\times$ \\
  tpmsgx ~\cite{tian2019a} & SGX & Container & Sealing & N/A & N/A  & N/A & $\times$ \\
  \textbf{SvTPM} & SGX & KVM/QEMU & Sealing & \checkmark & SGX $\rightarrow$ SvTPM  & \checkmark & \checkmark \\
  \bottomrule
  \end{tabular}}
\begin{tablenotes}
\footnotesize
\item* N/A: the feature is not considered or not explicitly elaborated in the work.
\end{tablenotes}
\label{tbl:table1}
\end{threeparttable}
\end{table*}

\section{Discussion}

{\bfseries Relationship between TPM and SGX.}  SvTPM aims at providing a secure virtual TPM using SGX in the cloud. There is a question about whether we still need TPMs when SGX can be used. In our opinion, TPM and SGX are complementary technologies. TPM with trusted computing technology can be used to measure the integrity of the whole computing environment from the underlying BIOS, operating systems to the upper applications. While SGX technology can build isolated containers so as to ensure the run-time security for key applications. If these two technologies are used together, the security of the real system can be greatly improved.

{\bfseries TCB of SvTPM.}  In SvTPM, the SGX mechanism and the key modules of vTPM, such as Libtpms 2.0 and NVRAM, works as TCB. However, we assume hypervisor or VMM is benign-but-vulnerable, because virtual machine management operations, such as creation, stop and destruction, are controlled by the hypervisor in the cloud. If we assume that hypervisor or VMM is completely untrusted, the management operations of VMs and vTPMs, such as creation, are completely untrusted. That is actually impractical in a real cloud platform.  Hence, we assume that hypervisor or VMM is benign-but-vulnerable in our SvTPM system.

{\bfseries Trust path between the SvTPM enclave and the VM.}  In SvTPM, although we are able to ensure that the TPM command is executed safely within the enclave and the credentials of vTPM can not be leaked, the execution results ultimately need to be sent to the upper layer VM from SvTPM enclave. Because the enclave does not cover the entire communication path, there is still a risk that the execution results can be tampered with after being delivered by enclave. This issue can be resolved through a secure channel, such as SSL channel, between the enclave and the VM. Meanwhile, this method may incur an additional performance overhead.

\section{Related Works}
\label{sec:related}

vTPM is a virtual TPM instance, which is used in a virtual environment (e.g. cloud computing platform), to build trust computing base. The most important work about virtual TPM is that Stefan Berger et al.~\cite{berger2006vtpm} from IBM design and implement a vTPM system on a XEN-based virtual platform. They virtualize TPM by extending the standard TPM command set to support vTPM life-cycle management and extend vTPM trust to a hardware TPM in the virtual environment. Frederic Stumpf et al.~\cite{stumpf2007} propose a scheme to build a trusted virtual platform. They bind vTPM with physical TPM, and construct of vTPM certificate chain. Their solution uses the AIK to sign virtual AIK (vAIK), physical PCR, nonce, and time-stamp, which are then used to form vAIK certificates. Paul England and Jork Loeser~\cite{paravtpm} extend the hypervisor with virtual PCR (vPCR) and TPM context manager. This approach allows guest operating systems to share hardware TPM. But the number of virtual machines on a physical machine is uncertain. Thus, this approach suffers from performance bottleneck due to the limited memory space of TPM. Matthew Fioravante and Daniel De Graaf~\cite{Matthew2012vTPM} present a virtual trusted platform module for Shielded VMs, which uses a light VM (i.e. MiniOS) to isolate vTPM in Xen platform. Every virtual machine has a dedicated VM as vTPM. Meanwhile, their work still uses a physical TPM to provide the identity trust for the vTPM.  Virtin Spector~\cite{VirtinSpector2014} propose to put vTPM to SMM mode so as to provide the isolation for the vTPM. However, this method has a higher performance overhead because entering SMM mode requires suspending all other CPU cores. cTPM ~\cite{chenctpm} presents an extension of the TPM's design that adds an additional root key to the TPM and shares that root key with the cloud. Therefore, cloud users can create and share TPM-protected keys and data across multiple devices they own.

The most related works to SvTPM are eTPM~\cite{sunetpm}, fTPM~\cite{rajftpm} and tpmsgx~\cite{tian2019a}. Their main difference are analyzed and depicted in Table 1.

eTPM ~\cite{sunetpm} designs a new trusted cloud platform security component enclave TPM to protect cloud. Both SvTPM and eTPM are built upon Intel SGX. However, they are different in several aspects: (1) eTPM relies on physical TPM chips, but SvTPM does not use TPM chips, improving runtime performance; and (2) eTPM does not present and solve the security challenges for vTPM protection, such as NVRAM replacement attacks, rollback attacks to vTPM and secure clock.

fTPM~\cite{rajftpm} presents a design of a firmware-based TPM, which leverages ARM TrustZone. They analyze security guarantees and shortcomings of the ARM TrustZone technology and then present three approaches to overcome the limitations of ARM TrustZone. fTPM is mainly for ARM-based mobile devices, not for vTPMs in cloud environments. Although fTPM also stores the TPM states on untrusted storage, its challenges and solutions are different compared with our solution, especially considering NVRAM replacement attacks and trust establishment between vTPM and SGX platform.

Dave Tian  et al.~\cite{tian2019a} presents lxcsgx, which allows SGX applications to run inside containers. This solution puts a software-based TPM to an SGX enclave. However, compared with their work, SvTPM presents and resolves a number of crucial security challenges when providing secure vTPMs for cloud tenants using SGX. In addition, SvTPM can support KVM/QEMU architecture, TPM 2.0 specification, and many cryptographic algorithms.

Heaven~\cite{baumann2015shielding} leverages SGX and Drawbridge, a sandbox, to provide an isolated PICO process container. It implements the shielded execution of unmodified legacy applications in an untrusted cloud. The system library and shield modules are put in SGX enclaves. The system library interacts with Drawbridge host through Downcalls and Upcalls to complete system functions required by user programs. Shield module is responsible for checking function parameters and returns results to ensure secure execution of user programs. SCONE~\cite{arnautov2016scone} proposes a secure Docker container based on SGX. It supports user-level threads and asynchronous system calls to reduce the high-performance overhead due to thread synchronization and system calls in the SGX enclave. Graphene-SGX~\cite{tsai2017graphene-sgx} implements a number of improvements to make security benefits of SGX more usable, such as the integrity support for dynamically-loaded libraries, and secure multi-process so as to rapidly deploy unmodified applications on SGX with small overheads. VC3~\cite{schuster2015vc3} presents a trusted computing environment for big data based on SGX technology to ensure the data security  of computation and storage. It seals the data and code of users, and then the cloud operating system loads the data and code into an isolated enclave. Furthermore, an efficient distributed job execution protocol is proposed to ensure the correctness and confidentiality of MapReduce jobs in all computing nodes.

The above works leverage SGX for building a secure computing environment for key applications. In contrast, SvTPM focuses on protecting the key security modules, vTPM, using SGX in the cloud.

\section{Conclusions}
In this paper, we highlight limitations of the state-of-the-art vTPM protection solutions, such as lacking run-time security protection and low performance, due to relying on the hardware TPM chip and weak isolation mechanism. We then propose a secure and efficient software-based vTPM, SvTPM, to address the challenges including NVRAM replacement attacks, rollback attacks, trust establishment between vTPMs and SGX platform, and the secure clock by using SGX to protect vTPMs. We implement an SvTPM prototype system and evaluate its performance. The evaluation results show that SvTPM has better performance and provides strong run-time isolation protection for vTPMs in the cloud.

\bibliographystyle{plain}
\bibliography{svtpm}

\begin{thebibliography}{10}

\bibitem{armbrust2010a}
M.~Armbrust, A.~Fox, R.~Griffith, A.~D. Joseph, R.~Katz, A.~Konwinski, G.~Lee,
  D.~Patterson, A.~Rabkin, I.~Stoica, and M.~Zaharia.
\newblock {A view of cloud computing}.
\newblock {\em Communications of the ACM}, 53(4):50--58, 2010.

\bibitem{arnautov2016scone}
S.~Arnautov, B.~Trach, F.~Gregor, T.~Knauth, A.~Martin, C.~Priebe, J.~Lind,
  D.~Muthukumaran, D.~O'keeffe, M.~Stillwell, D.~Goltzsche, D.~Eyers,
  K.~R{\"u}diger, and C.~Fetzer.
\newblock {SCONE: Secure Linux Containers with Intel SGX.}
\newblock In {\em OSDI}, volume~16, pages 689--703, 2016.

\bibitem{bade2010scalable}
Steven~A Bade, Charles~Douglas Ball, Ryan~Charles Catherman, James~Patrick
  Hoff, and James~Peter Ward.
\newblock Scalable paging of platform configuration registers, January~26 2010.
\newblock US Patent 7,653,819.

\bibitem{baumann2015shielding}
A.~Baumann, M.~Peinado, and G.~Hunt.
\newblock Shielding applications from an untrusted cloud with haven.
\newblock {\em ACM Transactions on Computer Systems (TOCS)}, 33(3):8, 2015.

\bibitem{berger2006vtpm}
S.~Berger, R.~Cáceres, K.~A. Goldman, R.~Perez, R.~Sailer, and L.~Van Doorn.
\newblock {vTPM: virtualizing the trusted platform module}.
\newblock In {\em Proc. 15th Conf. on USENIX Security Symposium}, pages
  305--320, 2006.

\bibitem{brenner2016securekeeper}
Stefan Brenner, Colin Wulf, David Goltzsche, Nico Weichbrodt, Matthias Lorenz,
  Christof Fetzer, Peter Pietzuch, and R{\"u}diger Kapitza.
\newblock Securekeeper: confidential zookeeper using intel sgx.
\newblock In {\em Proceedings of the 17th International Middleware Conference},
  page~14. ACM, 2016.

\bibitem{sgxpse}
S.~Cen and B.~Zhang.
\newblock {Trusted Time and Monotonic Counters with Intel Software Guard
  Extensions Platform Services}.
\newblock Intel Corporation White Paper, 2017.

\bibitem{chenctpm}
C.~Chen, H.~Raj, S.~Saroiu, and A.~Wolman.
\newblock {cTPM: A Cloud TPM for Cross-Device Trusted Applications.}
\newblock In {\em NSDI}, pages 187--201, 2014.

\bibitem{chen2018racing}
G.~Chen, W.~Wang, T.~Chen, S.~Chen, Y.~Zhang, X.~Wang, T.-H. Lai, and D.~Lin.
\newblock {Racing in hyperspace: closing hyper-threading side channels on SGX
  with contrived data races}.
\newblock In {\em Racing in Hyperspace: Closing Hyper-Threading Side Channels
  on SGX with Contrived Data Races}. IEEE, 2018.

\bibitem{Chen2017}
S.~Chen, X.~Zhang, M.~K. Reiter, and Y.~Zhang.
\newblock {Detecting privileged side-channel attacks in shielded execution with
  D{\'e}j{\'a} Vu}.
\newblock In {\em Proceedings of the 2017 ACM on Asia Conference on Computer
  and Communications Security}, pages 7--18. ACM, 2017.

\bibitem{costan2016intel}
V.~Costan and S.~Devadas.
\newblock {Intel SGX Explained.}
\newblock {\em IACR Cryptology ePrint Archive}, 2016(086):1--118, 2016.

\bibitem{cucurull2014virtual}
J.~Cucurull and S.~Guasch.
\newblock {Virtual TPM for a secure cloud: fallacy or reality?}
\newblock 2014.

\bibitem{dai2017rollsec}
W.~Dai, Y.~Du, H.~Jin, W.~Qiang, D.~Zou, S.~Xu, and Z.~Liu.
\newblock {RollSec: Automatically Secure Software States Against General
  Rollback}.
\newblock {\em International Journal of Parallel Programming}, pages 1--18,
  2017.

\bibitem{paravtpm}
P.~England and J.~Loeser.
\newblock {Para-Virtualized TPM Sharing}.
\newblock In {\em {Trusted Computing - Challenges and Applications}}, pages
  119--132. Springer Berlin Heidelberg, Berlin, Heidelberg, 2008.

\bibitem{stumpf2008enhancing}
Frederic F.~Stumpf and C.~Eckert.
\newblock Enhancing trusted platform modules with hardware-based virtualization
  techniques.
\newblock In {\em Emerging Security Information, Systems and Technologies,
  2008. SECURWARE'08. Second International Conference on}, pages 1--9. IEEE,
  2008.

\bibitem{Matthew2012vTPM}
Matthew Fioravante and Daniel~De Graaf.
\newblock {The virtual Trusted Platform Module (vTPM) subsystem for Xen}.
\newblock \url{https://xenbits.xen.org/docs/4.6-testing/misc/vtpm.txt}, Nov
  2012.

\bibitem{vmware}
M.~Foley.
\newblock {vSphere 6.7 – Virtual Trusted Platform Modules}.
\newblock
  \url{https://blogs.vmware.com/vsphere/2018/05/vsphere-6-7-virtual-trusted-platform-modules.html},
  May 2018.

\bibitem{fu2017sgx-lapd}
Y.~Fu, E.~Bauman, R.~Quinonez, and Z.~Lin.
\newblock {Sgx-Lapd: Thwarting Controlled Side Channel Attacks Via Enclave
  Verifiable Page Faults}.
\newblock In {\em International Symposium on Research in Attacks, Intrusions,
  and Defenses}, pages 357--380. Springer, 2017.

\bibitem{tpmlibrary}
Trusted~Computing Group.
\newblock {TPM Library Specification, Family "2.0", Level 00, Revision 01.38}.
\newblock
  \url{https://trustedcomputinggroup.org/resource/tpm-library-specification/}.

\bibitem{tpmmain}
Trusted~Computing Group.
\newblock {TPM Main Specifiction Level 2 Version 1.2, Revision 116}.
\newblock
  \url{https://trustedcomputinggroup.org/resource/tpm-main-specification/}.

\bibitem{tpmsummary}
Trusted~Computing Group.
\newblock {Trusted Platform Module(TPM) Summary}.
\newblock White Paper, Apr 2008.

\bibitem{gruss2017strong}
D.~Gruss, J.~Lettner, F.~Schuster, O.~Ohrimenko, I.~Haller, and M.~Costa.
\newblock Strong and efficient cache side-channel protection using hardware
  transactional memory.
\newblock In {\em USENIX Security Symposium}, pages 217--233, 2017.

\bibitem{han2018a}
Seunghun Han, Wook Shin, Junhyeok Park, and Hyoungchun Kim.
\newblock A bad dream: Subverting trusted platform module while you are
  sleeping.
\newblock In {\em 27th USENIX Security Symposium}, pages 1229--1246, 2018.

\bibitem{VirtinSpector2014}
Yan Fei. Shi Xiang. Li Zhihua. Wang Juan.~Zhang Huanguo.
\newblock Virtinspector: A uefi based dynamic secure measurement framework for
  virtual machine.
\newblock {\em Journal of Sichuan University(Engineering Science Edition)},
  46(1):22--28, 2014.

\bibitem{intelsgx}
Intel.
\newblock {Intel Software Guard Extensions}.
\newblock \url{https://software.intel.com/en-us/sgx/}.

\bibitem{sgxguide}
Intel.
\newblock {\em {Intel Software Guard Extensions Developer Guide}}, May 2018.

\bibitem{sgxreference}
Intel.
\newblock {\em {Intel Software Guard Extensions SDK for Linux* OS Developer
  Reference}}, September 2018.

\bibitem{epid}
S.~Johnson, S.~Vinnie, R.~Carlos, B.~Ernie, and M.~Frank.
\newblock {Intel Software Guard Extensions: EPID Provisioning and Attestation
  Service}.
\newblock
  \url{https://software.intel.com/sites/default/files/managed/57/0e/ww10-2016-sgx-provisioning-and-attestation-final.pdf},
  2016.

\bibitem{krautheim2010introducing}
F.~J. Krautheim, D.~S. Phatak, and A.~T. Sherman.
\newblock Introducing the trusted virtual environment module: a new mechanism
  for rooting trust in cloud computing.
\newblock In {\em International Conference on Trust and Trustworthy Computing},
  pages 211--227. Springer, 2010.

\bibitem{liu2010cloud}
D.~Liu, J.~Lee, J.~Jang, S.~Nepal, and J.~Zic.
\newblock A cloud architecture of virtual trusted platform modules.
\newblock In {\em 2010 IEEE/IFIP International Conference on Embedded and
  Ubiquitous Computing}, pages 804--811. IEEE, 2010.

\bibitem{lombardi2011secure}
F.~Lombardi and R.~Di Pietro.
\newblock Secure virtualization for cloud computing.
\newblock {\em Journal of network and computer applications}, 34(4):1113--1122,
  2011.

\bibitem{matetic2017rote}
S.~Matetic, M.~Ahmed, K.~Kostiainen, A.~Dhar, D.~Sommer, A.~Gervais, A.~Juels,
  and S.~Capkun.
\newblock {ROTE: Rollback Protection for Trusted Execution.}
\newblock {\em IACR Cryptology ePrint Archive}, 2017:48, 2017.

\bibitem{vanbulck2018foreshadow}
J.~Van Bulckand~M. Minkin, O.~Weisse, D.~Genkin, B.~Kasikci, F.~Piessens,
  M.~Silberstein, T.~F. Wenisch, Y.~Yarom, and R.~Strackx.
\newblock {Foreshadow: Extracting the Keys to the Intel SGX Kingdom with
  Transient Out-of-Order Execution}.
\newblock In {\em 27th USENIX Security Symposium}, pages 991--1008, 2018.

\bibitem{hyper-v}
C.~Nemnonm.
\newblock {How To Enable Virtual TPM(vTPM) in Windows Server 2016 Hyper-V}.
\newblock
  \url{https://charbelnemnom.com/2017/03/how-to-enable-virtual-tpm-vtpm-in-windows-server-2016-hyper-v-vm-hyperv-ws2016/},
  Apr 2018.

\bibitem{rajftpm}
H.~Raj, S.~Saroiu, A.~Wolman, R.~Aigner, J.~Cox, P.~England, C.~Fenner,
  K.~Kinshumann, J.~Loeser, D.~Mattoon, M.~Nystrom, D.~Robinson, R.~Spiger,
  S.~Thom, and D.~Wooten.
\newblock {fTPM: A Software-Only Implementation of a TPM Chip.}
\newblock In {\em USENIX Security Symposium}, pages 841--856, 2016.

\bibitem{schuster2015vc3}
F.~Schuster, M.~Costa, C.~Fournet, C.~Gkantsidis, M.~Peinado, G.~Mainar-Ruiz,
  and M.~Russinovich.
\newblock {VC3: Trustworthy data analytics in the cloud using SGX}.
\newblock In {\em Security and Privacy (SP), 2015 IEEE Symposium on}, pages
  38--54. IEEE, 2015.

\bibitem{shao2015formal}
J.~Shao, Y.~Qin, D.~Feng, and W.~Wang.
\newblock {Formal analysis of enhanced authorization in the TPM 2.0}.
\newblock In {\em Proceedings of the 10th ACM Symposium on Information,
  Computer and Communications Security}, pages 273--284. ACM, 2015.

\bibitem{shih2017t-sgx}
M.-W. Shih, S.~Lee, T.~Kim, and M.~Peinado.
\newblock {T-SGX: Eradicating controlled-channel attacks against enclave
  programs}.
\newblock In {\em Proceedings of the 2017 Annual Network and Distributed System
  Security Symposium (NDSS), San Diego, CA}, 2017.

\bibitem{shinde2016preventing}
S.~Shinde, Z.~L. Chua, V.~Narayanan, and P.~Saxena.
\newblock Preventing page faults from telling your secrets.
\newblock In {\em Proceedings of the 11th ACM on Asia Conference on Computer
  and Communications Security}, pages 317--328. ACM, 2016.

\bibitem{stumpf2007}
F.~Stumpf, M.~Benz, M.~Hermanowski, and C.~Eckert.
\newblock {An approach to a trustworthy system architecture using
  virtualization}.
\newblock In {\em International Conference on Autonomic and Trusted Computing},
  pages 191--202. Springer, 2007.

\bibitem{sunetpm}
H.~Sun, R.~He, Y.~Zhang, R.~Wang, W.~Ip, and K.~Yung.
\newblock {eTPM: A Trusted Cloud Platform Enclave TPM Scheme Based on Intel SGX
  Technology}.
\newblock {\em Sensors}, 18(11):3807, 2018.

\bibitem{tian2019a}
Dave Tian, Joseph~I Choi, Grant Hernandez, Patrick Traynor, and Kevin R~B
  Butler.
\newblock A practical intel sgx setting for linux containers in the cloud.
\newblock 2019.

\bibitem{tsai2017graphene-sgx}
C.-C. Tsai and D.~E. Porterand~M. Vij.
\newblock {Graphene-SGX: A practical library OS for unmodified applications on
  SGX}.
\newblock In {\em 2017 USENIX Annual Technical Conference (USENIX ATC)}, 2017.

\bibitem{whitaker2002denali}
A.~Whitaker, M.~Shaw, and S.~D. Gribble.
\newblock {Denali: Lightweight virtual machines for distributed and networked
  applications}.
\newblock Technical report, Technical Report 02-02-01, University of
  Washington, 2002.

\bibitem{xia2012defending}
Y.~Xia, Y.~Liu, H.~Chen, and B.~Zang.
\newblock {Defending against VM rollback attack}.
\newblock In {\em Dependable Systems and Networks Workshops (DSN-W), 2012
  IEEE/IFIP 42nd International Conference on}, pages 1--5. IEEE, 2012.

\bibitem{googlecloud}
M.~Zimmerman.
\newblock {Virtual Trusted Platform Module for Shielded VMs: security in
  plaintext}.
\newblock
  \url{https://cloud.google.com/blog/products/gcp/virtual-trusted-platform-module-for-shielded-vms-security-in-plaintext},
  Jul 2018.

\end{thebibliography}

\end{document}